\documentclass[preprint,pre,aps,onecolumn,superscriptaddress]{revtex4}
\usepackage{graphicx}
\usepackage{amsmath}
\usepackage{amsfonts}
\usepackage{amssymb}
\usepackage{multirow}
\usepackage{diagrams-v7}
\usepackage{float}
\usepackage{relsize}
\newcommand{\blangle}{\Big\langle}
\newcommand{\brangle}{\Big\rangle}

\begin{document}
\title{Active matter beyond mean-field: ring-kinetic theory for self-propelled particles}
\author{Yen-Liang Chou}
\affiliation{Max Planck Institute for the Physics of Complex Systems, N{\"o}thnitzer Stra{\ss}e 38, 01187 Dresden, Germany}
\author{Thomas Ihle}
\affiliation{Department of Physics, North Dakota State University, Fargo, North Dakota, 58108-6050}
\affiliation{Max Planck Institute for the Physics of Complex Systems, N{\"o}thnitzer Stra{\ss}e 38, 01187 Dresden, Germany}

\begin{abstract}
Recently, Hanke {\em et al.} [Phys.Rev. E {\bf 88}, 052309 (2013)] showed that mean-field kinetic theory fails to describe
collective motion in soft active colloids and that correlations must not be neglected. Correlation effects are also expected to be
essential in systems of biofilaments driven by molecular motors and in swarms of midges.
To obtain correlations in an active matter system from first principles, we derive
a ring-kinetic theory for Vicsek-style models of self-propelled agents
from the exact $N$-particle evolution equation in phase space.
The theory goes beyond mean-field and does not rely on Boltzmann's approximation of molecular chaos.
It can handle
pre-collisional correlations and cluster formation which both seem important to understand the phase transition to collective motion.
We propose a diagrammatic technique to perform a small density expansion of the collision operator 
and derive the first two equations of the BBGKY-hierarchy.
An algorithm is presented that numerically solves the evolution equation for the two-particle correlations on a lattice.
Agent-based simulations are performed and informative quantities such as orientational and density correlation functions are compared with those obtained 
by ring-kinetic theory.
Excellent quantitative agreement between simulations and theory is found
at not too small noises and mean free paths. 
This shows that there is parameter ranges in Vicsek-like models where the correlated closure 
of the BBGKY-hierarchy gives correct and nontrivial results.
We calculate the dependence of the orientational correlations on distance in the disordered phase
and find that 
it seems to be consistent with  
a power law with exponent around -1.8, followed by an exponential decay.
General limitations of the kinetic theory and its numerical solution are discussed.
\end{abstract}


\maketitle


\section{Introduction}

Recently, collective motion of active matter 
has been studied intensively in 
theories, simulations and experiments \cite{vicsek_12,ramaswamy_10,marchetti_13,toner_98}.    
In particular, great progress has been made in theoretical studies using kinetic theory approaches 
\cite{bussemaker_97,helbing_98,bertin_06,baskaran_08a,baskaran_08b,bertin_09,ihle_11,chou_12,peshkov_12,ihle_13,grossmann_13,thueroff_13,hanke_13,bertin_13, 
ihle_14_a,ihle_14_b,ihle_14_c, peshkov_14_a,peshkov_14_b,chepizhko_14}
which provide a bridge from microscopic dynamics to hydrodynamic equations.
The kinetic transport equations have been used to study the nature of the phase transition to collective motion, 
the stability of the ordered phase, and the morphology of emerging structures. 
Many of these studies focus on one of the simplest and most popular models of self-propelled particles -- 
the Vicsek-model (VM) \cite{vicsek_95,czirok_97,nagy_07} 
and its variants \cite{peruani_08,aldana_09,peng_09,barbaro_09,chou_12,ginelli_10a,romensky_14,mishra_12}. 
Due to the simplicity of its interaction rules that still lead to rich collective behavior, 
the VM became an archetype
of active matter.
Despite the minimality of the VM, its phase behavior is still not 
very well understood. Agent-based simulations at
large particle velocities show that the onset of collective motion is linked to the formation of  
high-density bands \cite{nagy_07,chate_08}. The bands are typically aligned with
the walls of the periodic simulation box and reach percolating size.

While it is known that these soliton-like bands can be quantitatively described by kinetic 
theory and provide a mean-field mechanism to render the flocking transition 
discontinuous \cite{ihle_13,caussin_14}, the situation at small particle velocities, where correlation effects are expected to be important, remains elusive.
In particular, in Ref. \cite{nagy_07} 
it was reported that bands are absent in this more physical regime of small mean free path. 
In addition, some researchers have interpreted band formation and the related discontinuous nature of the flocking transition
as numerical artifacts induced by periodic boundary conditions \cite{nagy_07,baglietto_09,baglietto_09b,aldana_09}. 
Other groups see band formation at the threshold to collective motion as inevitable, 
in the thermodynamic limit of the Vicsek-model
\cite{toner_priv,caussin_14,solon_14}.
Based on simulations of percolating bands at large mean free path, 
a reinterpretation of the flocking transition in terms of
a liquid-gas transition was recently proposed \cite{solon_13,solon_14}.
This description builds on hydrodynamic theories which are either phenomenological \cite{toner_98,caussin_14} or 
were obtained under mean-field assumptions \cite{bertin_06,bertin_09,ihle_11,baskaran_08a,mishra_10}, neglecting correlation effects.

In 2013, Hanke {\em et al.} \cite{hanke_13} adapted the collision kernel of the mean-field kinetic theory of Bertin {\em et al.} 
\cite{bertin_06,bertin_09} for  
{\em soft active colloids}. 
Their surprising result was that if orientational correlations were neglected, kinetic theory fails, that is, it 
predicts
the absence of collective motion which is clearly at odds with corresponding molecular dynamics simulations.
Such correlations are likely to be essential for other experimental systems as well \cite{schaller_10,sumino_12,weber_13}. 
Thus, there appears to be a need for
an analytical approach to active matter systems that includes correlations and calculates them 
from first principles.
In the particular case of the Vicsek-model, an approach is needed 
that remains valid at small mean free paths where correlations could impact
band formation.
Such a theory would deepen 
our understanding of
the ordering process in active systems and could lead to hydrodynamic equations with an extended range of validity.

The kinetic theory proposed for Vicsek-style models by directly adopting the Boltzmann equation \cite{bertin_06, bertin_09, peshkov_14_a} 
is based on two following assumptions. 
First, only binary collisions are assumed to occur.
This assumption is an intrinsic property of a Boltzmann-like kinetic theory. 
It was introduced
because the likelihood of genuine three- and more-particle collisions in a dilute, regular gas with short-ranged repulsion is small
compared to binary encounters.
The second, more serious, assumption is that the mean-free path is long enough 
for collisional partners to escape from each other and to loose 
the memory of their encounter before the next collision.
This is the molecular chaos assumption, originally called ``Stosszahl Ansatz'' by Boltzmann, which is usually 
reserved for the low density regime.
At high density, strongly correlated events, such as re-collisions, ring-collisions and cage-diffusion, become relevant 
\cite{dorfman_77,cohen_93a,cohen_93b,ernst_98}. 

On one hand, since the molecular chaos approximation is equivalent to a mean-field assumption it leads to a 
huge simplification of the corresponding kinetic theories, and became very popular. 
On the other hand, molecular chaos is not plausible for 
active and granular matter systems when the relative velocity between particles is greatly reduced after a collision and when
the mean free path is short.
This is especially true in systems with alignment interactions, such as the Vicsek-model 
near or in the phase
of collective motion. Here, particles form clusters and stay together for quite some time, repeatedly undergoing correlated 
collisions. 
Currently, 
an accurate bottom-up theory for 
the order/disorder transition 
of self-propelled particles 
with relevant cluster formation is lacking, although some progress has been made 
by means of a rate-equation approach \cite{peruani_06,peruani_10}.
The ring-kinetic approach explored here is able to quantitatively describe the effects of moderate 
clustering \cite{FOOT0}. Therefore, we hope that this paper will be useful 
on the way to a detailed theoretical understanding of the 
transition to collective motion. 

To get a first idea about the possible failure of the mean-field assumption one can 
compare its predictions for the transition to collective motion
with agent-based simulations. For the Vicsek-model at low densities and velocities, it is found that the 
theory overestimates the threshold noise by a factor 
between two and three \cite{FOOT0A}. 
More detailed critiques on the molecular chaos assumption in active matter can be found in Refs. \cite{chou_12,thueroff_13,ihle_14_proceeding}.
Recently, it was shown explicitly for the Vicsek-model (in the low speed regime and close to the flocking transition) 
that the binary collision assumption is also not valid, not even at very low particle densities
\cite{ihle_14_a}. 

A kinetic theory for Vicsek-like
models,
called phase-space or Enskog-like approach, 
was recently developed by 
one of us \cite{ihle_11,ihle_14_a}.
It is not restricted to low densities and binary collisions but
can handle collisions of an arbitrary number of partners \cite{FOOT1}.
Like most kinetic theories of active matter, it still assumes molecular chaos.
However, in this approach, molecular chaos is not an uncontrolled approximation. Instead, it is adjusted 
by an additional small parameter $\varepsilon=R/\lambda$, the ratio of the interaction radius $R$ to the mean free path $\lambda=\tau v_0$, where
$\tau$ is the finite time step and $v_0$ is the speed of particles in the Vicsek-model. For $\varepsilon \rightarrow 0$, 
molecular chaos becomes exactly valid \cite{FOOT2}. 
On the downside, in the VM at low densities, we only found good agreement between mean-field theory and agent-based simulations for unrealistically 
long mean free paths $\lambda$ of a least five times the radius of interaction $R$ \cite{chou_12}.
This is quite an unphysical regime because it allows agents to pass each other at very short 
distances 
without interaction.
Improving this unrealistic situation requires to go beyond mean-field and was a main motivation for this study.

Mathematically, the molecular chaos assumption is usually implemented by replacing N-particle distribution functions 
by products of one-particle functions. This leads to a non-correlated closure in the Boltzmann-like theory
and reduces the infinite 
BBGKY (Bogoliubov-Born-Green-Kirkwood-Yvon) hierarchy of equations \cite{kirkwood_35,bogo_46,hansen_book} to just the first equation.
Recently, Hanke {\em et al.} \cite{hanke_13} 
have tried to ``repair'' the first BBGKY-equation by including correlation effects obtained from 
agent-based scattering simulations, see also \cite{thueroff_PRX}.
Recent extensions of Dynamic Density Functional Theory \cite{marconi_99,marconi_00,wittkowski_11} 
to active systems \cite{wensink_08,speck_14,menzel_14} 
also contain 
correlation effects in an approximated form by assuming that certain functional relations 
known from
equilibrium systems are still valid out of equilibrium.
In Ref. \cite{bialke_13} a 
Smoluchowski approach for self-propelled repulsive disks was approximately closed by introducing a 
force coefficient which is proportional to an integral over pair-correlations
but remains an undetermined parameter. 
To the best of our knowledge, nobody has attempted yet to self-consistently account for correlation and memory effects 
in Vicsek-style models 
by closing a BBGKY-like hierarchy at a higher level and explicitly solving the second hierarchy equation. 
The second equation describes the time evolution of the two-particle correlation function and has the potential to predict long-ranged
positional and orientational correlations.
Such an approach is called ring-kinetic theory and has led to many interesting results in regular and granular fluids 
such as the calculation of
the so-called long-time tails and long-ranged spatial correlations 
\cite{dorfman_77,cohen_93a,cohen_93b,ernst_98,dorfman_70,pomeau_71,mazenko_73,mehaffey_77,kirkpatrick_91,brito_92,ernst_95,noije_98}.
In this paper, we take the first step beyond the mean-field assumption of molecular chaos for self-propelled particles. 
We set up the so-called repeated-ring kinetic theory for a Vicsek-style model  
and solve the second BBGKY-like equation numerically in the limit of small density.
In the long-term, we aim to answer the following more fundamental question: Is it possible to set-up a first-principle
theory that quantitatively describes far-from-equilibrium systems of many interacting objects even in parameter ranges
where mean-field theories fail?  

In repeated-ring kinetic theory, both the one-particle density $f_1$ and the two-particle density $f_2$ provide input to the temporal evolution of $f_1$ and $f_2$, whereas higher order correlations are neglected.
This allows the implicit treatment of correlated interaction sequences, called ring-collisions.
To give an example of a ring-collision, consider three initially uncorrelated particles and assume that particle 1 first
interacts with particle 2, then particle 2 interacts with particle 3.
Finally, assume that an instant later, particle 1 collides with particle 3.
Even though particles 1 and 3 have never met directly, their interaction has
pre-collisional correlations because they were in contact with the same particle 2 in the past, 
and as a result, carry information about their common experience with particle 2.

Ring-kinetic theory is tedious and has significant limitations, 
which probably contributed to its rather low popularity after the 1970s \cite{dorfman_67,ernst_69,ernst_98}. %
In our case, the difficulty level forced us to develop diagrammatic representations of collision integrals.
In addition, to arrive at analytically solvable integrals for the many different coupling constants, we 
slightly modified the collision rule of the standard Vicsek-model. Instead of the original multi-particle alignment
rule we use binary collisions where the focal particle randomly picks a single collision partner from the ones available in
a circle of radius $R$ around its position.
At low densities and in the absence of strong clustering, this rule becomes identical to the one of the standard VM.

A more serious issue of ring-kinetic theory is that it still needs a closure-condition to truncate the BBGKY-hierarchy.
The traditional closure consists of setting all connected n-particle correlations with $n\ge 3$ to zero. 
This is reasonable in regular gases at low density but the validity of this truncation is far from obvious in systems of active matter. 

In the current approach, we still use this traditional closure but control it in the same way 
as we managed the molecular chaos assumption in the mean-field version of the phase-space approach:
We know that for $\varepsilon=R/(v_0 \tau)\rightarrow 0$ molecular chaos becomes exact and 
all connected correlation functions should become negligible.
It seems plausible that there is a range of small but nonzero $\varepsilon$ where the two-particle correlations dominate the 
three-particle and higher n-particle correlations. 
This hypothesis can be justified {\em a posteriori} 
through quantitative agreement between  
ring-kinetic theory and agent-based simulations, something we indeed find at not too large $\varepsilon$.
Direct measurements of three- and four-particle correlations in agent-based simulations confirm the existence of such a ``weak-coupling''-regime 
and will be reported elsewhere
\cite{ihle_14_proceeding}

The main results of this paper are (i) the construction of the repeated-ring kinetic 
theory of a Vicsek-style model that includes pre-collisional 
correlations and thus goes beyond mean-field, 
(ii) the introduction of a diagrammatic expansion of the collision operator in powers of the density, and 
(ii) the demonstration of excellent quantitative agreement of the theoretical predictions for the orientational and positional correlations
with agent-based simulations at sufficiently large noise and mean free paths.
We also provide data to explicitly show the limitations of our current approach, which seems to require a more sophisticated closure
when the noise is very small, and both density and mean free path are also small.

The paper is organized as follows:
In Section ~\ref{sec:model} we introduce the modified Vicsek-model, which we will call {\em binary Vicsek-model} (BVM), and
derive the first two BBGKY-like hierarchy equations for the VM and BVM in Section ~\ref{sec:RING}.
In addition, the rules for the diagrammatic expansion of the collision operator are introduced
and motivated in this section.
The algorithm to solve the hierarchy equations is explained in Section ~\ref{sec:numerics}.
In Section ~\ref{sec:results} the results of the numerical evaluation of these kinetic equations are presented and compared
to agent-based simulations. A summary is given in Section ~\ref{sec:conclusion}.
Details concerning the evaluation of coupling integrals
are relegated to Appendix A.
In Appendix B, a
list of diagrams for the second BBGKY-hierarchy equation can be found. 
In Appendix C, we explore parameter regions in which discrepancies between the current kinetic theory and microscopic simulations occur.

\section{Microscopic model}
\label{sec:model}

The standard Vicsek-model consists of 
$N$ point particles with mean number density $\rho_0$. 
The particles with
positions ${\bf x}_i(t)$ and velocities ${\bf v}_i(t)=v_0({\rm cos}(\theta_i), {\rm sin}(\theta_i))$
undergo discrete-time dynamics
with time step $\tau$. 
The velocities are uniquely characterized by the 
flight direction $\theta_i$ because the particles move in two dimensions at the same constant speed $v_0$. 
In the so-called streaming step
all positions are updated according to
\begin{equation}
\label{VM_UPDATE}
{\bf x}_i(t+\tau)={\bf x}_i(t)+\tau {\bf v}_i(t)\,.
\end{equation}
In the subsequent collision step,
particles align with their neighbours within a fixed distance $R$ by updating their flight directions.
In particular,
a circle of radius $R$ is drawn around a given particle and the average direction $\Phi_i$ of motion
of all 
particles
within the circle is determined
according to
\begin{equation}
\Phi_i={\rm arctan}[\sum_{\{j\}} {\rm sin}(\theta_j)/\sum_j^n {\rm cos}(\theta_j)]\,,
\end{equation}
Then, the new particle directions
are determined as
\begin{equation}
\label{ANGLE_RULE}
\theta_i(t+\tau)=\Phi_i+\xi_i
\end{equation}
 where $\xi_i$ is a random number which is
uniformly distributed in
the interval $[-\eta/2,\eta/2]$.
Note, that the updated positions ${\bf x}_i(t+\tau)$ (and not the old locations ${\bf x}_i(t)$)
are used to determine the average directions $\Phi_i$.
The updates are parallel and correspond to the so-called forward updating rule, 
see Refs. \cite{huepe_08,baglietto_09}.

Although the kinetic formalism of Section \ref{sec:RING} does apply to the standard VM,
a slightly modified version of the standard algorithm is used in our practical implementations. 
In this version, which we will label {\em binary Vicek model} (BVM), the calculation of the average direction $\Phi_i$ contains
additional randomness: Instead of including all particles found in a circle around the focal particle $i$ into the calculation, 
only {\em one} collision partner  
is selected  with equal probability $1/(n-1)$, given that there are 
$n-1$ potential collision partners inside the circle.
At very low local densities, most circles 
will only contain the focal particle, that is $n=1$, or one additional particle corresponding to $n=2$.
In this case, the binary VM is identical to the standard VM.
The motivation for introducing the BVM is two-fold.
First, it provides a huge technical advantage in ring-kinetic theory because the coupling integrals, 
defined in 
Eqs. (\ref{eq.coupling_definition}), can be solved analytically for arbitrary particle numbers $n$.
For the standard VM, only the cases $n=1,2$ and the asymptotic situation $n\rightarrow \infty$ appear to be analytically 
solvable. Therefore, one would have to rely
on large tables of numerically calculated integrals.

The second motivation for a microscopic model with random but strictly binary interactions comes from dense systems
of granular and active particles with volume exclusion. 
In these systems, particles rattle around in cages formed by their neighbors \cite{cohen_93b}.
But even if the density is quite high, 
particles will mostly be in contact with only one or two others at a given instant 
because of their very short ranged interaction. 
However, the frequency of these encounters will increase with density. 
The binary VM tries to emulate this scenario in a very crude way: 
it replaces genuine multi-particle collision by a stochastic sequence of
binary encounters. Of course, in true caging, the sequence of collision partners is correlated while it is not in BVM.
Nethertheless, some aspects of systems with higher densities should be captured by this new model.

\begin{figure}
\begin{center}
\includegraphics[width=3.1in,angle=0]{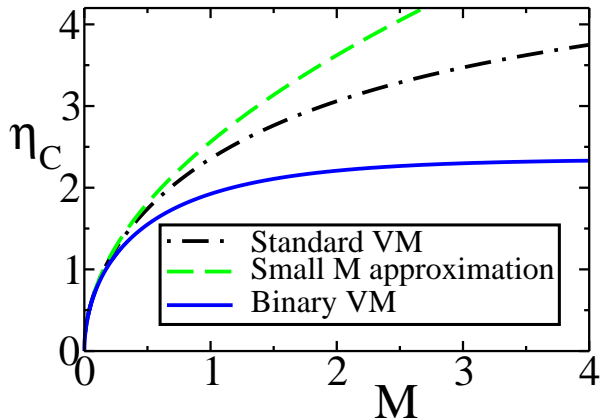}
\caption
{
The mean-field phase diagram of a homogeneous system for the binary VM (blue solid line) obtained from Eq. 
(\ref{PHASE_BVM})
in comparison to the standard VM (dashed-dotted line) and 
the small density approximation, Eq. (\ref{BVM_CRIT}), (green dashed line).
Noise values below a particular line, that is at $\eta<\eta_C$, correspond to global order.
Note that in systems larger than a critical linear size $L_C$, inhomogeneous, wave-like states occur that would alter
the phase diagram plotted here \cite{ihle_11,ihle_13}.
}
\label{FIG_PHASE}
\end{center}
\end{figure}

An additional technical advantage of BVM is that the mean-field phase diagram for a homogeneous system, that is the dependence 
of the threshold noise $\eta_C$ on the normalized density $M=\pi R^2\rho_0$, can be calculated analytically for all densities.
The inverse relation $M(\eta_C)$ is given by
\begin{eqnarray} 
\nonumber
M&=&-{\rm ln}\left[{\gamma-(4/\pi) \over 1-(4/\pi)}\right]\;\;{\rm with} \\
\gamma&=&{\eta_C\over 2\, {\rm sin}(\eta_C/2)}
\label{PHASE_BVM}
\end{eqnarray}
and shown in Fig. \ref{FIG_PHASE}.
Note, that for the standard VM, analytical results can only be obtained asymptotically for small and high $M$, see Refs. 
\cite{ihle_11,ihle_14_a}, such as,
\begin{equation}
\label{BVM_CRIT}
\eta_C=\sqrt{48 M\left({2\over \pi}-{1\over 2}\right)}\;\;\;{\rm for}\;M\ll 1
\end{equation}
As expected, expanding Eq. (\ref{PHASE_BVM}) for $M\ll 1$ reproduces the results of the standard VM, Eq. (\ref{BVM_CRIT}), 
see Fig. \ref{FIG_PHASE}.
The biggest difference in the phase diagrams occur in the infinite density limit, $M\rightarrow \infty$. In this limit, the critical 
noise for BVM does not reach the largest possible angle of $2\pi$ as in the standard VM \cite{FOOT6}.
Instead, one obtains the maximum critical noise $\eta_{\infty}\approx 2.345$ from the transcendental equation
\begin{equation}
\label{BVM_TRANS}
\pi={8\over \eta_{\infty}} {\rm sin}\left({\eta_{\infty}\over 2}\right)\,.
\end{equation}
Using agent-based simulations we have checked 
that phenomena known from the standard VM such as the formation of spiky soliton-like density waves 
\cite{bertin_09,aldana_09,ihle_13} also occur in BVM.

A side effect of the BVM collision rule is that interactions can become directional. For example, let us assume that the mutual distances 
between three particles is  less than the interaction radius $R$. 
Now, particle $1$ could pick particle $2$ to align with but at the same time, particle $2$ might choose to 
ignore $1$ and aligns with particle $3$ instead.
This cannot occur in the standard VM: 
Particle $1$ has to include particle $2$ in determining its new direction, and reciprocally,
particle $2$
will include particle $1$ in its interaction.
This subtle difference leads to more interaction possibilities and to more terms in the diagrammatic expansion, which is 
discussed in Appendix A.

\section{Ring-kinetic theory}
\label{sec:RING}

\subsection{Derivation of the BBGKY hierarchy}
\label{sec:BBGKY}

The microscopic state of a Vicsek-like model at a given time $t$ is 
fully described by the $N$-particle probability density function 
$P_N(Z_1,Z_2,\cdots,Z_N,t)$, characterizing the probability of finding particles in the infinitesimal phase space volume $dZ_1dZ_2\cdots dZ_N$ around the phase $(Z_1,Z_2,\cdots,Z_N)$. 
Here, $Z_i\equiv(\mathbf{X}_i, \mathbf{V}_i)$ marks the position and velocity of the $i$-th particle. 
Since the particle speed in the VM is assumed to be constant and equal to $v_0$, 
one usually uses the polar representation $(V, \Theta_i)$ of $\mathbf{V}_i$ or simply the orientation $\Theta_i$ instead, to 
describe the motion of the particle. 
In this paper we will alternatively use $\mathbf{V}_i$ and $\Theta_i$ without specifying.

The general form of the evolution equation for the $N$-particle distribution function, that describes a Markov chain in phase space, was 
given by Ihle \cite{ihle_11,ihle_14_a}
\begin{equation}
P_N(Z'_1,Z'_2,\cdots,Z'_N,t+\tau)={\mathbb C}_N\circ P_N(Z_1,Z_2,\cdots,Z_N,t)~,
\label{eq.master_equation_for_N_particles}
\end{equation}
Here, $Z'_i=(\mathbf{X}'_i, \mathbf{V}'_i)=(\mathbf{X}'_i, \Theta_i')$ is the new coordinate of the $i$-th 
particle after one iteration of the collision and streaming processes. 
The collision operator ${\mathbb C}_N$ takes the form 
\begin{equation}
{\mathbb C}_N
 ={1\over \eta^N} \prod_{i=1}^N\int_{-\eta/2}^{\eta/2}\, d\xi_i\,
\int_0^{2\pi} {d\Theta_i}\,
\hat{\delta}(\Theta_i'-\Phi_i-\xi_i)~,
\label{eq.collsion_operator_for_N_particles}
\end{equation}
where $\Phi_i$ is the mean direction of the particles inside the collisional zone of the $i$-th particle, $\xi_i$ the angular noise added to the aligned orientation $\Phi_i$ bounded in the interval $[-\frac{\eta}{2},\frac{\eta}{2}]$. 
The kernel of the collision operator consists of products of the periodic Dirac delta function 
$\hat{\delta}(x)=\delta(x~\text{modulo}~2\pi)$. This delta function 
gives the transition rate of the $i$-th particle from its pre-collisional angle $\Theta_i$ to the post-collisional angle $\Theta_i'$, which is non-zero only if the condition, 
$\Theta_i'=\Phi_i+\xi_i$, is satisfied. 
To account for all ways to create a specific post-collisional state, 
integrations
over
the pre-collisional angles $\Theta_i$ and over the angular noises $\xi_i$ must be performed.
We note that the new velocities $\mathbf{V}'_i$ are updated via the collisional operator ${\mathbb C}_N$ 
while the new positions are obtained through the streaming $\mathbf{X}'_i=\mathbf{X}_i+\tau\mathbf{V}'_i$
which is implicitly denoted by the argument $Z'_i$ on the left hand side of the kinetic equation (\ref{eq.master_equation_for_N_particles}).

The full description by Eqs. (\ref{eq.master_equation_for_N_particles}, \ref{eq.collsion_operator_for_N_particles}), is exact but contains too much information for practical application.
The standard way to proceed \cite{ernst_81,hansen_book} is
to first derive a reduced $S$-particle probability distribution function (PDF) by integrating the full PDF over the coordinates $Z_{S+1},Z_{S+2},\cdots,Z_N$
\begin{equation}
P_S(Z_1,Z_2,\cdots,Z_S)
=\int P_N(Z_1,Z_2,\cdots,Z_N)\, dZ_{S+1} dZ_{S+2} \cdots dZ_{N}\,, 
\label{eq.reducing_integral_Pn}
\end{equation}
to obtain a reduced $S$-particle kinetic equation.
Usually, the reduced $S$-particle equation relates the $S$-particle- to the $(S+1)$-particle PDF. 
The full set of the reduced equations, which contains the same information as the original evolution equation, 
is called the BBGKY (Bogoliubov-Born-Green-Kirkwood-Yvon) hierarchy, see for example Refs. 
\cite{kirkwood_35,bogo_46,hansen_book}. 

The hierarchy equations become useful if 
the macroscopic properties can be well described already by the averages taken with respect to the first 
few reduced PDF's instead of the full description. 
In general, this assumption constitutes a big leap of faith but in our case
the results of Section \ref{sec:results} show that there is parameter ranges in the VM where this
is justified.
Here, we derive the first two equations of the BBGKY hierarchy for the reduced one- and
two-particle densities $f_1$ and $f_2$. This is done by evaluating the ensemble average 
of their microscopic counterparts, namely
\begin{eqnarray}
f_1(z_1)&=&\int dZ^{(N)} P_N(Z_1,Z_2,\cdots,Z_N)\, \Psi_1(z_1)
\label{eq.reducing_integral_f1}
\\
f_2(z_1,z_2)&=&\int dZ^{(N)} P_N(Z_1,Z_2,\cdots,Z_N)\, \Psi_2(z_1,z_2)
\label{eq.reducing_integral_f2}
\end{eqnarray}
where $dZ^{(N)}$ is short for $dZ_1 dZ_2\cdots dZ_N$, and $z_i\equiv(\mathbf{x}_i,\mathbf{v}_i)\equiv(\mathbf{x}_i,\theta_i)$ denote 
field variables which have to be distinguished from the particle phases $Z_i$.
For brevity, we have omitted the time-dependence of $f_j$, $P_N$, $Z_i$, and $\Psi_j$ in our notation.
The microscopic one-particle density is defined as
\begin{equation}
\label{PSI1}
\Psi_1(z_1)=\sum_{i=1}^N\delta(Z_i-z_1),
\end{equation}
and simply gives the time-dependent density of particles in the three-dimensional $\mu$-space of the VM.
It is only non-zero if at a given time $t$ at 
least one particle happens to be at the specified field point $z_1\equiv (x_1,y_1,\Theta_1)$.

Similarily, the microscopic two-particle density, see for example Ref. \cite{klimon_67_74},
\begin{equation}
\label{PSI2}
\Psi_2(z_1,z_2)=\sum_{i=1}^N\sum_{j\neq i}^N \delta(Z_i-z_1)\delta(Z_j-z_2),
\end{equation}
accounts for simultaneously finding one particle at $z_1$ and another at $z_2$, where $\delta(Z_i-z_j)\equiv\delta(\mathbf{X}_i-\mathbf{x}_j)\delta(\Theta_i-\theta_j)$.
The one-particle density $f_1$ is normalized to the number of particles $N$, while the two-particle function $f_2$ is normalized to the number of ordered pairs, $N(N-1)$. 
This is different from the probability distribution function $P_S$ which is normalized to unity for any $S$. 
Inserting Eqs. (\ref{PSI1}, \ref{PSI2}) into (\ref{eq.reducing_integral_f1}, \ref{eq.reducing_integral_f2}) 
and using definition (\ref{eq.reducing_integral_Pn}), the following relations are obtained, 
\begin{eqnarray}
\label{eq.P1_to_f1}
f_1(z_1)&=&N P_1(z_1)~,\\
\label{eq.P2_to_f2}
f_2(z_1,z_2)&=&N(N-1) P_2(z_1,z_2)~.
\end{eqnarray}

To facilitate the derivation of the hierarchy equations from the full evolution equation, Eq.  
(\ref{eq.master_equation_for_N_particles}), 
we expand the $N$-particle distribution function by means of the Ursell expansion which is also known as 
cluster expansion, see for example Refs. \cite{ernst_81,kubo_62,onuki_78}.  
The Ursell expansion is a set of hierarchical expansions 
in terms of the so-called connected correlation functions $G_S$. 
These functions account for the excess information beyond the product distribution and possess the so-called cluster property: Assume a system without long-ranged correlations
and consider a group of $n$ particles that are located very close to each other. 
If a single one of these particles 
is moved away from the others, $G_n$ for these particles will go to zero, whereas $P_n$ would not.
The first two expansions are shown below
\begin{eqnarray}
\nonumber
P_2(Z_1,Z_2)&=&P_1(Z_1)P_1(Z_2)+G_2(Z_1,Z_2)~,\\\nonumber
P_3(Z_1,Z_2,Z_3)&=& P_1(Z_1)P_1(Z_2)P_1(Z_3)\\\nonumber
                &+& P_1(Z_1)G_2(Z_2,Z_3)    \\\nonumber
                &+& P_1(Z_2)G_2(Z_3,Z_1)    \\\nonumber
                &+& P_1(Z_3)G_2(Z_1,Z_2)    \\
\label{Ursell}
                &+& G_3(Z_1,Z_2,Z_3)~.
\end{eqnarray}
Accordingly, a full expansion for the $N$-particle distribution function can be written down.
Important conditions on $G_S$ follow from the marginalization of $P_S$ to $P_{S-1}$, Eq. (\ref{eq.reducing_integral_Pn}),
\begin{equation}
\label{eq.marginalization_condition}
\int_{\mbox{all}}d{\bf X}_i \int_0^{2\pi} d\Theta_i\,G_S(Z_1,Z_2,\cdots,Z_S)=0\,,\;\;\;\;\;\;\;i=1,2\ldots S
\end{equation}
where the subscript ``all'' refers to a spatial integration over the entire volume.
We will call this relation ``normalization condition'' because if it is violated, the $N$-particle
probability density $P_N$ would not be normalized to unity anymore.

In the following, we will also need spatial integration of a particle position over the collision area which is either a circle or
a union of two circles. This integration is denoted by the subsript ``{\rm in}''. 
The 
complementary
operation, which consists of an integration over all space {\em except} the collision area, is labeled by the subscript ``out''. 
This gives,
\begin{equation}
\label{SPLIT1}
\int_{\mbox{all}}d\mathbf{X}_i\ldots=
\int_{\mbox{out}}d\mathbf{X}_i\ldots+\int_{\mbox{in}}d\mathbf{X}_i\ldots
\end{equation}
This integral splitting and Eq. (\ref{eq.marginalization_condition}) allow us to rewrite the integration over the outside region 
as an integration over the collision zone,
\begin{equation}
\label{IN_OUT_TRICK}
\int_{\mbox{out}} d{\bf X}_i \int_0^{2\pi} d\Theta_i \,G_S(Z_1,Z_2,\cdots,Z_S)=
-\int_{\mbox{in}} d{\bf X}_i \int_0^{2\pi} d\Theta_i \,G_S(Z_1,Z_2,\cdots,Z_S)
\end{equation}
which will lead to significant advantages in solving the BBGKY-equations.

Finally, in analogy to the relation between $P_2$ and $f_2$, see Eq. (\ref{eq.P2_to_f2}), we introduce 
a rescaled two-particle correlation function $g_2$,
\begin{equation}
\label{gDEF}
g_2(z_1,z_2)=N(N-1) G_2(z_1,z_2)~.
\end{equation}
This leads to,
\begin{equation}
\label{f2_to_g}
f_2(z_1,z_2)=\left(1-{1\over N}\right)f_1(z_1) f_1(z_2)+g_2(z_1,z_2)~.
\end{equation}
For a system with finite particle number $N$ and vanishing correlations, $g_2=0$, one sees that $f_2$ is not exactly equal to the product of two $f_1's$. 
This feature is inherited from the definition of the two-particle density $\Psi_2$, Eq. (\ref{PSI2}), which assumes that 
the same particle cannot
simultaneously be found at two different locations ${\bf x}_1$ and ${\bf x}_2$.

To derive the reduced hierarchy equations for $f_1$ and $f_2$, we first plug the Ursell expansion 
into the right-hand side of the $N$-particle evolution equation, Eq. (\ref{eq.master_equation_for_N_particles}).
Then, we multiply both sides with the microscopic one- and two-particle density, respectively, and perform the marginalization procedure (\ref{eq.reducing_integral_f1}, \ref{eq.reducing_integral_f2}), 
\begin{eqnarray}
\label{MARGIN1}
f_1(\mathbf{x'}_1,\theta'_1,t+\tau)&=&\int d\mathbf{X}^{(N)}\,d\Theta'^{(N)} \Psi_1(\mathbf{x}_1,\theta'_1)\, 
~{\mathbb C}_N\circ P_N(\mathbf{X}^{(N)},\Theta^{(N)})~,\\
\label{MARGIN2}
f_2(\mathbf{x'}_1,\theta'_1,\mathbf{x'}_2,\theta'_2,t+\tau)&=&\int d\mathbf{X}^{(N)}\,d\Theta'^{(N)} 
 \Psi_2(\mathbf{x}_1,\theta'_1,\mathbf{x}_2,\theta'_2)\,
~{\mathbb C}_N\circ P_N(\mathbf{X}^{(N)},\Theta^{(N)})\,,
\end{eqnarray}
to obtain kinetic equations that do not depend on the particle phases but on field variables instead.
Here, the phases $(\mathbf{X}^{(N)},\Theta^{(N)})$ and the densities $\Psi_j$ on the right hand side are evaluated at time $t$.
We also have
$\mathbf{x'}_i=\mathbf{x}_i+\tau \mathbf{v'}_i$ with $\mathbf{v'}_i=v_0({\rm cos}\theta_i',{\rm sin}\theta_i')$.

\subsection{Diagrammatic approach}
\label{sec:diagram}

To illustrate how the integrations in the first two hierarchy equations, Eqs. (\ref{MARGIN1},\ref{MARGIN2}), 
can be simplified in a systematic manner, let us consider a specific term in the Ursell expansion of a 
$10$-particle system, namely $P_1(Z_1)P_1(Z_2)G_2(Z_3,Z_4)G_2(Z_5,Z_6)G_2(Z_7,Z_8)P_1(Z_9)P_1(Z_{10})$ that occurs in the right
hand side of Eq. (\ref{MARGIN1}).
This term describes three pairs of particles that are correlated through two-particle correlations.
The rest of the particles is uncorrelated.
The outcome of a collision will depend on where these particles are located with respect to each other.
For example, if the distance between particles 3 and 4 is smaller than the radius $R$ of the collision circle and all 
other particle are far away from them, a correlated collision between 3 and 4 will occur. 
Since the collision integral, Eq. (\ref{MARGIN1}) involves an integration over all particle positions, the above situation 
is just one of the
many possible collision scenarios that have to be considered.
The main idea to evaluate collision integrals of this kind is to first classify all possibilities and then
to integrate over just one member of each class. The other members, which give the same contribution, are incorporated
by combinatorial prefactors.  

The microscopic density $\Psi_1$ is defined at one focal point, $\mathbf{x}_1$, whereas the two-particle density $\Psi_2$
depends on two focal points, $\mathbf{x}_1$ and $\mathbf{x}_2$. The delta functions in the definition of $\Psi_1$ 
together with the integration of the particle positions lead to terms in Eq. (\ref{MARGIN1}) 
where one particular particle $i$, $i=1,2,\ldots N$,
called the focal particle, is fixed at $\mathbf{x}_1$.
Analogously, in the second equation (\ref{MARGIN2}), we have two focal particles. 
In this mathematical formalism, one focal particle has to ``stay'' at $\mathbf{x}_1$ and 
the other is forced to ``stay'' at $\mathbf{x}_2$.
Of course, since all particles are identical, it does not matter which ones are the focal ones and we just choose 
particle $1$ to be the focal particle in Eq. (\ref{MARGIN1}), and particles $1$ and $2$ to be the focal particles in the second hierarchy
equation. The other choices lead to combinatorial factors of $N$ and $N(N-1)$, respectively.

Once the focal particles are chosen, we have to classify the situation with respect to the locations of the remaining particles.
For the first BBGKY-equation, Eq. (\ref{MARGIN1}), these classes are defined by how many of the uncorrelated particles
are located inside the collision circle around the focal particle, how many correlated pairs are inside this circle and how
many correlated pairs have one member of the pair outside the circle and the other one inside.
For the second hierarchy equation, the situation is more complicated, since the collision scenario will also depend
on the distance between the two focal particles.
As shown below in Eq. (\ref{EXAMPLE_P10}), such a classification is much easier to handle in terms of diagrams. 
In our example for Eq. ({\ref{MARGIN1}), 
we assume there is one uncorrelated particle (labeled 2) located in the circle around particle 1. 
We further assume that there is one correlated 
pair (consisting of particles 3 and 4) inside and one pair (particles 7 and 8) outside the circle. 
The remaining pair has one particle inside and one particle outside the circle.  
The remaining degrees of freedom for this scenario are the specific positions of particles $1,2,\ldots 5$ {\em within} the collision circle 
and the specific postions of the particles $6,7,\ldots 10$ {\em outside} the circle.
This means that in the spatial integrations, the first group of particles, $1,2,\ldots 5$, is not allowed to ``leave'' the collision circle, whereas the latter
group has to ``stay'' outside. Scenarios which violate this rule are not neglected but 
either belong to different
diagrams or to different members of the same class.

It is straightforward to write down the contribution from the term above to the evolution of the $1$-particle density $f_1$,
\begin{eqnarray}
\label{EXAMPLE_P10}
\BigTenA[3] &=& \frac{10!}{8}
\int \frac{d\xi}{\eta}
\int_{\mbox{in}} d\mathbf{X}_2 d\mathbf{X}_3\cdots d\mathbf{X}_5
\int_{\mbox{out}}d\mathbf{X}_6 d\mathbf{X}_7\cdots d\mathbf{X}_{10}
\\\nonumber
&& \int d\Theta_1 d\Theta_2 \cdots d\Theta_{10} \hat{\delta}[\theta'_1-\xi-\Phi_1(\Theta_1,\Theta_2,\cdots\Theta_5)]
\\\nonumber
&& P_1(\mathbf{x}_1,\Theta_1)P_1(\mathbf{X}_2,\Theta_2)
P_1(\mathbf{X}_9,\Theta_9)P_1(\mathbf{X}_{10},\Theta_{10})
\\\nonumber
&& G_2(\mathbf{X}_3,\Theta_3,\mathbf{X}_4,\Theta_4)
G_2(\mathbf{X}_5,\Theta_5,\mathbf{X}_6,\Theta_6)
G_2(\mathbf{X}_7,\Theta_7,\mathbf{X}_8,\Theta_8)
\end{eqnarray}
On the left-hand side, we use a diagram to represent this complicated equation.
We use "$\ssf$" to denote the focal particle at the selected position ${\bf x}_1$. 
Here, this selected particle is uncorrelated and is represented by the 
$1$-particle distribution function $P_1({\bf x}_1,\Theta_1)$. 
The symbols "$\ssF$" stand for independent particles that lead to factors of $P_1(Z_j)$, 
while the correlated particle pairs are represented by the link "$\ssX$" that stands for the connected 
correlation function $G_2(Z_i,Z_j)$. The big circle which encloses particles inside the collision zone of the focal particle 
represents angular, spatial and noise integrations under 
the restriction that particles are not allowed to cross the circumference of the circle. 
The numbers in the diagram are particle labels. They indicate just one possible realization 
of a particular class and are given for reference. 

We are interested in the total number of ways to form a specific diagram. In this case, the combinatorial factor is $10!/8$. 
The integration of an independent particle outside the circle yields $1-\frac{M(\mathbf{x}_1)}{N}$, where
$M$ is the local average particle number in the circle centered around ${\bf x}_1$, $M=\int_{in}\rho({\bf x})\,d{\bf x}$.
According to Eq. (\ref{IN_OUT_TRICK}), the integration of a correlated particle outside the circle can be translateded into an integral over the 
inside of the circle with a negative sign.
However, these transfer-particles are ``virtual'' in the sense that they must not participate 
in the collision process of the focal particle and need to be distinguished from the genuine inside-particles. 
We use the unfilled circle "$\ssO$" to denote these particles and arrive at the following simplification,
\begin{equation}
\BigTenA[3]=\Bigg(1-\frac{M(\mathbf{x}_1)}{N}\Bigg)^2\BigTenB[3],
\end{equation}
with
\begin{eqnarray}
\label{DIAGRAMM_EX2}
\BigTenB[3] &=& -\frac{10!}{8} \frac{1}{N^2}\frac{1}{\left(N(N-1)\right)^3}\int \frac{d\xi}{\eta} 
\int_{\mbox{in}}d\mathbf{x}_2 d\mathbf{x}_3\cdots d\mathbf{x}_8
\\\nonumber
&&\int d\theta_1 d\theta_2 \cdots d\theta_8 \hat{\delta}[\theta'_1-\xi-\Phi_1(\theta_1,\theta_2,\cdots\theta_5)]
\\\nonumber
&& f_1(\mathbf{x}_1,\theta_1)f_1(\mathbf{x}_2,\theta_2)
g_2(\mathbf{x}_3,\theta_3,\mathbf{x}_4,\theta_4)
\\\nonumber
&&g_2(\mathbf{x}_5,\theta_5,\mathbf{x}_6,\theta_6)
g_2(\mathbf{x}_7,\theta_7,\mathbf{x}_8,\theta_8)
\end{eqnarray}
The negative sign appears because we have ``brought'' a total of three correlated particles to the inside of the circle. 
We have furthermore replaced $P_1$ by $f_1/N$ and $P_2$ by $f_2/(N(N-1))$ and change the particle's variable $(\mathbf{X}_i,\Theta_i)$ to the field variable $(\mathbf{x}_i,\theta_i)$. Note, the combinatorial factor is easier to count in this modified diagram. Here, we choose 
eight out of ten particles to form the diagram and there are 3 pairs but only one is an ordered pair. 

Having these diagrammatic representations and neglecting three-particle and higher correlations, we can write down the first hierarchy equation
for $N \to \infty$:
\begin{equation}
\label{eq.1st_BBGKY_full}
f_1(\mathbf{x}'_1,\theta'_1,t+\tau)=e^{-M(\mathbf{x}_1)}~\mathlarger{\mathlarger{\sum}}_{p,q,r,s}~\mathlarger{\Bigg\{}~
\ManyPA[0.8] + \ManyPB[0.8] + \ManyPC[0.8] ~\mathlarger{\Bigg\}}\,, 
\end{equation}
where
$\mathbf{x'}_1=\mathbf{x}_1+\tau \mathbf{v'}_1$.
The summation goes over $p$ dots, $q$ solid-solid, $r$ solid-open and $s$ open-open dumbbells in each sub-diagrams on the right-hand side of the equation, where $p$, $q$, $r$, and $s$ are integers running from $0$ to $\infty$.
The factor $e^{-M(\mathbf{x}_1)}$ comes from the contribution of infinitely many independent particles outside the circle according to the
limit,
\begin{equation}
\lim_{N\to \infty}\left(1-{M\over N}\right)^N={\rm e}^{-M}
\end{equation}

The $N$-dependent prefactor in the diagram (\ref{DIAGRAMM_EX2}) is compensated by aditional factors of $N$ and $N-1$ 
from the left side of the hierarchy equations
as well as from additional combinatorial factors due to the different choices of focal particles.
In the limit $N\rightarrow \infty$ and $M/N\rightarrow 0$ these factors converge to unity. 
Thus, the diagrams used in Eq. (\ref{eq.1st_BBGKY_full}) and all following equations look like the diagram of Eq. (\ref{DIAGRAMM_EX2}) but {\em without} the $N$-dependent prefactor.
Accordingly, in this limit, the particle number $N$ does not occur anymore in Eq. (\ref{eq.1st_BBGKY_full}).

Similarly, the second BBGKY equation can be constructed:
\begin{eqnarray}
\label{eq.2nd_BBGKY_full}
f_2(\mathbf{x}'_1,\theta'_1,\mathbf{x}'_2,\theta'_2,t+\tau)
&=&e^{-M_{12}(\mathbf{x}_1,\mathbf{x}_2)}~\sum_{p,q,r,s}~\bigg\{~\xUff + \xUg + \xUGf + \xUfG
\\\nonumber
&+&\xUHf + \xUfH + \xUGG + \xUHH + \xUGH + \xUHG~\bigg\}~,
\\
\label{nochmal_f2_to_g}
g_2(\mathbf{x}'_1,\theta'_1,\mathbf{x}'_2,\theta'_2,t+\tau)
&=&f_2(\mathbf{x}'_1,\theta'_1,\mathbf{x}'_2,\theta'_2,t+\tau)
\\\nonumber
&-&\left(1-\frac{1}{N}\right)f_1(\mathbf{x}'_1,\theta'_1,t+\tau)f_1(\mathbf{x}'_2,\theta'_2,t+\tau)~, 
\end{eqnarray}
where $M_{12}(\mathbf{x}_1,\mathbf{x}_2)$ is the average number of particles inside the union collision zone of $\mathbf{x}_1$ and $\mathbf{x}_2$.
The second relation (\ref{nochmal_f2_to_g}) follows from Eq. (\ref{f2_to_g}).
The shaded diagram is a simplified notation which implicitly contains $p$ dots, $q$ solid-solid, $r$ solid-open and $s$ open-open dumbbells, for example 
\begin{equation}
\xUff~\equiv\ManyPU[0.8].
\end{equation}

The symbol \ssu denotes integration over the union of two collisional circles. 
In this notation, the left and right crosses "$\ssf$" have coordinates $z_1$ and $z_2$, respectively, and correspond to the two 
focal particles.
For each particle to be integrated, the spatial domain of integration is divided into sub-regions depending on the distance between 
the two focal particles, $d$, for example,
\begin{equation}
\sUffF \equiv
\begin{cases}
~~ \sDaffFi +\sDaffFj +\sDaffFk  & ~\text{for } d \leq R\\
~~ \sDbffFi +\sDbffFj +\sDbffFk  & ~\text{for } R \leq d < 2R \\
~~ \sDcffFi +\sDcffFj  & ~\text{for } 2R < d~. 
\end{cases}
\end{equation}
Particles are not allowed to cross the boundaries of the sub-domains because this might change the outcome
of the collision step and would lead to double counting of the same proccess.
We summarize the notations used in the diagrammatic representation as following. The symbols "$\ssf$", "$\ssF$", and "$\ssO$" denote 
particles. A "link" between particles stands for a binary correlation between them. 
The symbols "$\ssc$" and "$\ssu$" are collisional operators which enclose particles involving in the collisional processes. The mathematical representations are listed in Table \ref{ta.diagrammatic_representation}. Note that in the current stage we only consider two-particle correlation functions, 
which works well for weakly-correlated systems. 
In Appendix C, we will discuss parameter regions of the VM where correlations beyond the binary ones cannot 
be neglected anymore.
\begin{table}
\def\arraystretch{1}
\begin{tabular}{|c|c|}
\hline
 symbol &   function/operator \\
\hline
\ssf \ssF      & $P_1(z_i) =\frac{1}{N}f_1(z_i)$   \\
\ssg \ssG \ssH & \multirow{2}{*}{$\pm G_2(z_i,z_j)=\pm\frac{1}{N(N-1)}g_2(z_i,z_j)$}   \\
\ssX \ssY \ssZ &   \\
\ssC           & 
$\displaystyle c_1\int \frac{d\xi}{\eta}\int_{\mbox{in}} d\mathbf{x}^{(n-1)}
\int d\theta^{(n)} \hat{\delta}\left[\theta'_1-\xi-\Phi_1\right] $ \\
\ssU           &
$\displaystyle c_2\int \frac{d\xi_1 d\xi_2}{\eta^2}\int_{\mbox{in}} d\mathbf{x}^{(n-2)}
\int d\theta^{(n)} \hat{\delta}\left[\theta'_1-\xi_1-\Phi_1\right] \hat{\delta}\left[\theta'_2-\xi_2-\Phi_2\right]$ \\
\hline

\end{tabular}
\caption{Notations used in the diagrammatic representation. The $\pm$ sign is given by $(-1)^k$ where $k$ is the number of the open circles $\ssO$. 
The combinatorial factors $c_1$ and $c_2$ count the number of ways to form the specific diagrams.}
\label{ta.diagrammatic_representation}
\end{table}

\subsection{Low density approximation and Fourier expansion}
\label{sec.Fourier_expansion}

In this section, 
we perform a small density expansion of the BBGKY equations. 
This is based on the assumption that the likelihood to find more than a few particles
in a collision circle is small when the average density $\rho_0=N/V$ is low.
In addition, we use Fourier expansions of the distribution functions with respect to their angular variables.
This allows us to integrate out the noise and the pre-collisional angles in the collision operators.
Let $f'_1(\mathbf{x}_1,\theta'_1)$ and $g'(\mathbf{x}_1,\theta'_1,\mathbf{x}_2,\theta'_2)$
be the density functions after collision but before streaming. 

For the small density expansion we use the dimensionless number $M$, 
that is, the average number of particles in a collision circle, as 
small expansion parameter. In the collision integral, products of $f$ and $g_2$ are multiplied by the $\hat{\delta}$-kernel 
and
are integrated over
the collision area.
Since such an integral over a single $f$ gives $M$ according 
to $M=\int d\theta \int_{circle}f({\bf x},\theta)\,d{\bf x}$
we assume that every factor of $f$ contributes a power of $M$ when counting the weight of a diagram.

Dimensional analysis of Eq. (\ref{f2_to_g}) reveals that $g_2$ has units of $f^2$. This suggests that 
every factor of $g_2$ in the collision integral contributes two powers of $M$. 
In terms of diagrams, this means
that each symbol which stands for a particle ($\ssf$,$\ssF$, and $\ssO$) carries one order of $M$. 
Thus, a diagram formed by $n$ particles is assumed to be of order $M^n$. 
For example, one has $\mathcal{O}(\sCfF[0.75])\sim M^2$ and $\mathcal{O}(\sUHf[0.75])\sim M^3$.
This naive way of judging the order of a diagram is intuitively appealing because in the low density limit 
where $M\ll 1$
it will be more likely to find just one particle in a circle than two or three.
Thus, for example, the diagram $\sCf$ will be considered more relevant than $\sCfFF$.
To obtain a consistent expansion in powers of $M$,
we also have to expand the exponential prefactors, such as $e^{-M}\approx 1-M+M^2/2+\ldots$.

For $N\rightarrow \infty$, the expansion of the first two BBGKY equations to order $M^2$ yields
\begin{equation}
f'_1(\mathbf{x}_1,\theta'_1)
= \left(1-M\right)\sCf+\sCfF+\sCG+\sCH~~,
\label{eq.1st_BBGKY_M2_expansion}
\end{equation}
and
\begin{equation}
g'_2(\mathbf{x}_1,\theta'_1,\mathbf{x}_2,\theta'_2)
= \sUff + \sUg -\sCf \times \sCf ~~,
\end{equation}
where the last term comes from the expansion of $f_1(\mathbf{x}_1,\theta_1) f_1(\mathbf{x}_2,\theta_2)$ to order $M^2$. 
In this and the following equations, whenever there is a multiplication of two diagrams, we asign the coordinate $z_1$ to the selected particle of the left diagram, and $z_2$ to the right.  

Similarly, expanding up to order ${\cal O}(M^3)$ gives, 
\begin{eqnarray}
\label{INFINITE_N_FIRST}
f'_1(\mathbf{x}_1,\theta'_1)
&=&\left(1-M+\frac{M^2}{2}\right)\sCf\\\nonumber
&+&(1-M) \Big(\sCfF+\sCG+\sCH\Big)\\\nonumber
&+&\sCfFF +\sCGF +\sCHF +\sCfX +\sCfY +\sCfZ ~~,
\end{eqnarray}
and
\begin{eqnarray}
\label{INFINITE_N_SECOND}
g'_2(\mathbf{x}_1,\theta'_1,\mathbf{x}_2,\theta'_2)
&=& (1-M_{12})\Big(\sUff + \sUg \Big)\\\nonumber 
&+& \sUffF + \sUgF \\\nonumber
&+& \sUGf  + \sUfG \\\nonumber 
&+& \sUHf  + \sUfH \\\nonumber
&-& (1-M_1-M_2)\Big(\sCf \times \sCf \Big) \\\nonumber
&-& \sCf \times \sCfF - \sCfF \times \sCf \\\nonumber
&-& \sCf \times \sCG  - \sCG  \times \sCf \\\nonumber
&-& \sCf \times \sCH  - \sCH  \times \sCf  ~~.
\end{eqnarray}

For small $N$, one has to use $(1-M/N)^{N-n}$ instead of $e^{-M}$ as the coefficient of the $n$-particle diagram, 
and similarly $(1-M_{12}/N)^{N-n}$ instead of $e^{-M_{12}}$ for the second equation.
For example, one replaces $1-M$ by $1-M/2$ in Eq.(\ref{eq.1st_BBGKY_M2_expansion}) for the $2$-particle system. 
For this special case of $N=2$, the resulting two hierarchy equations become exact, 
because no more particles are available to build higher order diagrams.
For $N>2$, the expansions to the order of $M^3$ are
\begin{eqnarray}
\label{FINITE_N_BBGKY}
f'_1(\mathbf{x}_1,\theta'_1)
&=&\left[1-\frac{N-1}{N}M+\frac{(N-1)(N-2)}{2N^2}M^2\right]~\sCf\\\nonumber
&+&\left(1-\frac{N-2}{N}M\right)~\Big(\sCfF+\sCG+\sCH\Big)\\\nonumber
&+&\sCfFF +\sCGF +\sCHF +\sCfX +\sCfY +\sCfZ ~~,
\end{eqnarray}
and
\begin{eqnarray}
\label{FINITE_N_SECOND}
g'_2(\mathbf{x}_1,\theta'_1,\mathbf{x}_2,\theta'_2)
&=& \left(1-\frac{N-2}{N}M_{12}\right)\Big(\sUff + \sUg \Big)\\\nonumber 
&+& \sUffF + \sUgF \\\nonumber
&+& \sUGf  + \sUfG \\\nonumber 
&+& \sUHf  + \sUfH \\\nonumber
&-& \left[1-\frac{N-2}{N}(M_1+M_2)\right]\Big(\sCf \times \sCf \Big) \\\nonumber
&-& \sCf \times \sCfF - \sCfF \times \sCf \\\nonumber
&-& \sCf \times \sCG  - \sCG  \times \sCf \\\nonumber
&-& \sCf \times \sCH  - \sCH  \times \sCf  ~~.
\end{eqnarray}

Our naive recipe of power counting does not take streaming into account, which presumably weakens 
three-particle
correlations more than two-particle correlations.
Note that the current way we assign powers of $M$ to diagrams implies that three- and four-particle correlations
would contribute at orders $O(M^3)$ and $O(M^4)$, respectively. 
Since these correlations are omitted in our current approach, we do not expect to gain 
much by expanding 
to orders higher than $O(M^3)$. 
Therefore, for particle numbers $N\ge3$ the equations (\ref{FINITE_N_BBGKY},\ref{FINITE_N_SECOND}) 
should be considered as weak-correlation approximations which assume that 
two-particle correlations dominate three-particle and higher correlations.
The consistency of these expansions with respect to conservation laws will be discussed in Section \ref{sec.conservation_laws}.

The Fourier expansions of the post-collisional functions are
\begin{eqnarray}
f'_1(\mathbf{x}_1,\theta'_1)&=&\sum_{m}\hat{f}'_m(\mathbf{x}_1)~e^{i m \theta'_1}~~,
\\
\label{FOURIER_LABEL}
g'_2(\mathbf{x}_1,\theta'_1,\mathbf{x}_2,\theta'_2)&=&
\sum_{m,n}\hat{g}'_{mn}(\mathbf{x}_1,\mathbf{x}_2)~e^{i m \theta'_1}e^{i n\theta'_2}~~.
\nonumber
\end{eqnarray}
where the Fourier modes are defined as, 
\begin{eqnarray}
\hat{f}'_m(\mathbf{x}_1)&=&\frac{1}{2\pi}\int_{-\pi}^{\pi}d\theta'_1~f'(\mathbf{x}_1,\theta'_1)~e^{-i m \theta'_1}~~,
\\\nonumber
\hat{g}'_{mn}(\mathbf{x}_1,\mathbf{x}_2)&=&\frac{1}{(2\pi)^2}\int_{-\pi}^{\pi}d\theta'_1 d\theta'_2~g'(\mathbf{x}_1,\theta'_1,\mathbf{x}_2,\theta'_2)~e^{-i m \theta'_1}e^{-i n\theta'_2}~~.
\end{eqnarray}
It is convenient to introduce the following notation for Fourier transformations,
\begin{eqnarray}
\blangle \cdots \brangle_m &\equiv &\frac{1}{2\pi}\int_{-\pi}^{\pi}d\theta'_1~\cdots~e^{-im\theta'_1}
\\
\blangle \cdots \brangle_{mn} &\equiv &\frac{1}{(2\pi)^2}\int_{-\pi}^{\pi}d\theta'_1 d\theta'_2~\cdots~e^{-i m \theta'_1}e^{-i n\theta'_2}~~
\nonumber 
\end{eqnarray}
Incorporating the collisional operators denoted by $\ssc$ and $\ssu$ one finds, 
\begin{eqnarray}
\label{eq.Fourier_mode_general}
\blangle \ssC \brangle_m &=&\frac{\lambda_m}{2\pi}\int d\theta^{(k)}\int d \mathbf{x}^{(k-1)}~\cdots~e^{-im\Phi_1}
\\
\blangle \ssU \brangle_{mn} &=&\frac{\lambda_{mn}}{(2\pi)^2}\int d\theta^{(k)} \int d\mathbf{x}^{(k-2)}~\cdots~e^{-i m \Phi_1}e^{-i n\Phi_2}~~,
\nonumber 
\end{eqnarray}
where $d\theta^{(k)}=\prod_{i=1}^k d\theta_i$ and $d\mathbf{x}^{(k-j)}=\prod_{i=j+1}^k d\mathbf{x}_i$ with $k$ being the number of particles enclosed by the collisional operator. 
The coefficients that result from integrating over post-collision angle(s) and the noise(s) are given by  $\lambda_m=\frac{2}{m\eta}\sin(\frac{m\eta}{2})$ for $m>0$, $\lambda_0=1$, and $\lambda_{mn}=\lambda_m\lambda_n$.
We also expand the pre-collisional distribution functions 
into series with coefficients $\hat{f}_p$ or $\hat{g}_{pq}$.
Inserting these expansions into the collision integrals, Eqs. (\ref{eq.Fourier_mode_general}), 
the integrations over the pre-collisional angles can be carried out and lead to the 
following coupling integrals,
\begin{eqnarray}
\label{eq.coupling_definition}
\mathrm{k}_{mpq}&=&\frac{1}{(2\pi)^2}
\int d\theta_1 d\theta_2
~e^{-im\Phi_1(\theta_1,\theta_2)}~e^{ip\theta_1}~e^{iq\theta_2}
\\\nonumber
\mathrm{k}_{mpqr}&=&\frac{1}{(2\pi)^3}
\int d\theta_1 d\theta_2 d\theta_3
~e^{-im\Phi_1(\theta_1,\theta_2,\theta_3)}~e^{ip\theta_1}~e^{iq\theta_2}~e^{ir\theta_3}
\\\nonumber
\mathrm{j}_{mnpq}&=&\frac{1}{(2\pi)^2}
\int d\theta_1 d\theta_2 
~e^{-im\Phi_1(\theta_1,\theta_2)}~e^{-in\Phi_2(\theta_1,\theta_2)}~e^{ip\theta_1}~e^{iq\theta_2}
\\\nonumber
\mathrm{i}_{mnpqr}&=&\frac{1}{(2\pi)^3}
\int d\theta_1 d\theta_2 d\theta_3
~e^{-im\Phi_1(\theta_1,\theta_3)}~e^{-in\Phi_2(\theta_2,\theta_3)}~e^{ip\theta_1}~e^{iq\theta_2}~e^{ir\theta_3}
\\\nonumber
\mathrm{h}_{mnpqr}&=&\frac{1}{(2\pi)^3}
\int d\theta_1 d\theta_2 d\theta_3
~e^{-im\Phi_1(\theta_1,\theta_2,\theta_3)}~e^{-in\Phi_2(\theta_1,\theta_2,\theta_3)}~e^{ip\theta_1}~e^{iq\theta_2}~e^{ir\theta_3}
\\\nonumber
\mathrm{l}_{mnpqr}&=&\frac{1}{(2\pi)^3}
\int d\theta_1 d\theta_2 d\theta_3
~e^{-im\Phi_1(\theta_1,\theta_2,\theta_3)}~e^{-in\Phi_2(\theta_1,\theta_2)}~e^{ip\theta_1}~e^{iq\theta_2}~e^{ir\theta_3}
\end{eqnarray}
At first sight, the dependence of the average angles $\Phi_i$ on up to three pre-collisional angles $\theta_1$, $\theta_2$ and $\theta_3$ in 
Eqs. (\ref{eq.coupling_definition}) seems to imply that these definitions apply merely to the standard Vicsek model and not
to the binary Vicsek model (BVM). This is because
in the BVM, only a maximum of two pre-collisonal angles directly contribute to the average angle.
In Appendix A we explain that this notation is to be interpreted as a symbolic notation and 
specify
how it can be translated such that it applies to both standard and binary VM.

Using the coupling constants from Eq. (\ref{eq.coupling_definition})
significantly simplifies 
the post-collisional terms.
For example,
\begin{equation}
\blangle \sCfF \brangle_m =
\frac{N(N-1)}{N^2} 
2\pi\lambda_m\sum_{pq}\mathrm{k}_{mpq}
\hat{f}_p(\mathbf{x}_1)
\int_{O_1} d\mathbf{x}_2 \hat{f}_q(\mathbf{x}_2),
\end{equation}
where $O_1$, the domain of the integration, is the area of the collision circle centered around $\mathbf{x}_1$ with radius $R$.
We will also frequently encounter the following special integrals. 
First, terms are needed, which involve an integration over the area inside the collision circle,
\begin{equation}
\bar{F}_m(\mathbf{x}_1)
\equiv \int_{O_1} d\mathbf{x}'_1\hat{f}_m(\mathbf{x}'_1).
\end{equation}
We also encounter cases where Fourier coefficients are integrated over the intersect of two circles centered around $\mathbf{x}_1$ and $\mathbf{x}_2$ separately. We denote this integral as
\begin{equation}
\Delta\bar{F}_m(\mathbf{x}_1,\mathbf{x}_2)
\equiv\int_{O_1\cap O_2} d\mathbf{x}'_1\hat{f}_m(\mathbf{x}'_1)=\Delta\bar{F}_m(\mathbf{x}_2,\mathbf{x}_1).
\end{equation}
Therefore the integration over the area $O_1$ but without $O_2$ (that takes the shape of a half-moon) is 
\begin{equation}
\int_{O_1\setminus O_2} d\mathbf{x}'_1\hat{f}_m(\mathbf{x}'_1)
=\bar{F}_m(\mathbf{x}_1)-\Delta\bar{F}_m(\mathbf{x}_1,\mathbf{x}_2)~.
\end{equation}
Second, regarding integrals that involve the two-particle correlation function, 
we define the first argument to be fixed at position $\mathbf{x}_1$, that is $\mathbf{x}'_1=\mathbf{x}_1$ but integrate the second 
argument $\mathbf{x}'_2$ over the circle centered around $\mathbf{x}_2$ as
\begin{equation}
\label{FIRST_BAR}
\bar{G}_{mn}(\mathbf{x}_1,\mathbf{x}_2)
\equiv\int_{O_2}d\mathbf{x}'_2\hat{g}_{mn}(\mathbf{x}_1,\mathbf{x}'_2)~,
\end{equation}
and over the intersection of the two circles
\begin{equation}
\label{SECOND_BAR}
\Delta\bar{G}_{mn}(\mathbf{x}_1,\mathbf{x}_2)
\equiv\int_{O_1\cap O_2}d\mathbf{x}'_2\hat{g}_{mn}(\mathbf{x}_1,\mathbf{x}'_2)~.
\end{equation}
Note that by definition $\Delta\bar{G}_{mn}(\mathbf{x}_1,\mathbf{x}_2)\neq\Delta\bar{G}_{mn}(\mathbf{x}_2,\mathbf{x}_1)$. 
This differs from $\Delta\Bar{F}_m(\mathbf{x}_1,\mathbf{x}_2)$ where the symmetry of interchangeing the variables $\mathbf{x}_1$ and $\mathbf{x}_2$ exists.
With the above definitions the following expressions can be derived:
\begin{eqnarray}
&&\int_{O_1}d\mathbf{x}'_2\,\hat{g}_{mn}(\mathbf{x}_1,\mathbf{x}'_2)
=\bar{G}_{mn}(\mathbf{x}_1,\mathbf{x}_1)
\\
&&\int_{O_2\setminus O_1}d\mathbf{x}'_2\,\hat{g}_{mn}(\mathbf{x}_1,\mathbf{x}'_2)
=\bar{G}_{mn}(\mathbf{x}_1,\mathbf{x}_2)-\Delta\bar{G}_{mn}(\mathbf{x}_1,\mathbf{x}_2)
\\
&&\int_{O_1\setminus O_2}d\mathbf{x}'_2\,\hat{g}_{mn}(\mathbf{x}_1,\mathbf{x}'_2)
=\bar{G}_{mn}(\mathbf{x}_1,\mathbf{x}_1)-\Delta\bar{G}_{mn}(\mathbf{x}_1,\mathbf{x}_2)
\end{eqnarray}
Last, we define the integration of both the variables over $O_1$
\begin{equation}
\label{THIRD_BAR}
\bar{\bar{G}}_{mn}(\mathbf{x}_1)
\equiv\int_{O_1}d\mathbf{x}'_1\int_{O_1}d\mathbf{x}'_2\,\hat{g}_{mn}(\mathbf{x}'_1,\mathbf{x}'_2)~.
\end{equation}
With all integrations defined, we give now a full list of the post-collisional Fourier 
modes for the individual diagrams up to order $\mathcal{O}(M^3)$.
For brevity, we only list the equations in the limit of $N\to\infty$. For small $N$ on has to restore the combinatorial and 
normalization factors, see eqs. (\ref{eq.P1_to_f1}, \ref{eq.P2_to_f2}). 
The Fourier modes for the first BBGKY-equation are given in digrammatic form as
\begin{eqnarray}
\blangle \sCf \brangle_m &=& 
\lambda_m \hat{f}_m(\mathbf{x}_1)
\\
\blangle \sCfF \brangle_m &=& 
2\pi\lambda_m\sum_{pq}\mathrm{k}_{mpq}
\hat{f}_p(\mathbf{x}_1)
\bar{F}_q(\mathbf{x}_1)
\\
\blangle \sCG \brangle_m &=& 
2\pi\lambda_m\sum_{pq}\mathrm{k}_{mpq}
\bar{G}_{pq}(\mathbf{x}_1,\mathbf{x}_1)
\\
\blangle \sCH \brangle_m &=& 
-2\pi\lambda_m\bar{G}_{m0}(\mathbf{x}_1,\mathbf{x}_1)
\\\nonumber
\\
\blangle \sCfFF \brangle_m &=& 
\frac{1}{2}(2\pi)^2\lambda_m\sum_{pqr}\mathrm{k}_{mpqr}
\hat{f}_p(\mathbf{x}_1)
\bar{F}_q(\mathbf{x}_1)
\bar{F}_r(\mathbf{x}_1)
\\
\blangle \sCGF \brangle_m &=& 
(2\pi)^2\lambda_m\sum_{pqr}\mathrm{k}_{mpqr}
\bar{G}_{pq}(\mathbf{x}_1,\mathbf{x}_1)
\bar{F}_r(\mathbf{x}_1)
\\
\blangle \sCfX \brangle_m &=& 
\frac{1}{2} (2\pi)^2\lambda_m\sum_{pqr}\mathrm{k}_{mpqr}
~\hat{f}_p(\mathbf{x}_1)
~\bar{\bar{G}}_{qr}(\mathbf{x}_1)
\\
\blangle \sCHF \brangle_m &=& 
-(2\pi)^2\lambda_m\sum_{pq}\mathrm{k}_{mpq}
\bar{G}_{p0}(\mathbf{x}_1,\mathbf{x}_1)
\bar{F}_q(\mathbf{x}_1)
\\
\blangle \sCfY \brangle_m &=& 
-(2\pi)^2\lambda_m\sum_{pq}\mathrm{k}_{mpq}
\hat{f}_p(\mathbf{x}_1)
\bar{\bar{G}}_{q0}(\mathbf{x}_1)
\\
\blangle \sCfZ \brangle_m &=& 
\frac{1}{2}(2\pi)^2\lambda_m 
\hat{f}_m(\mathbf{x}_1)
\bar{\bar{G}}_{00}(\mathbf{x}_1)
\end{eqnarray}  
To obtain the Fourier modes for the second BBGKY-equation,
three cases must be distinguished. For the {\em strong overlap} case with $d=|{\bf x}_2-{\bf x}_1|\le R$, 
the focal particles are within each others collision 
circle.
For example, $\blangle\sDaffFi\brangle_{mn}$ is a diagram for strong overlap.
The subscripts $m$ and $n$ denote Fourier labels related to the post-collisional angles.

The {\em weak overlap} scenario with $R< d \le 2R$ occurs if the focal particles cannot collide directly but 
could simultaneously interact with a third
particle that is located between them.
Finally, for $d > 2R$ there is {\em no overlap} of the two collision circles.
The digrams $\blangle\sDbGkf\brangle_{mn}$ and $\blangle\sDcfGi\brangle_{mn}$ are examples for weak and no overlap diagrams, respectively.
A full list of all relevant diagrams for the second BBGKY-equation up to order $O(M^3)$ is given in Appendix B.

\subsection{Physical quantities}
\label{sec.physical_quantities}

In this section we relate relevant physical observables to the Fourier modes of the density distributions.
In Section ~\ref{sec:results}, these relations will be used 
to compare kinetic theory predictions with agent-based simulation.
First, we consider the local number density at $\mathbf{x}$, which by definition is the average of the one-particle microscopic density integrated over the angular variable $\theta$ 
\begin{eqnarray}
\langle\rho(\mathbf{x})\rangle
&\equiv& \int d\mathbf{X}^{(N)} \int d\Theta^{(N)} P_N(\mathbf{X}^{(N)},\Theta^{(N)}) \int \Psi_1(\mathbf{x},\theta)\,d\theta
\\\nonumber
&=& N\int d\theta P_1(\mathbf{x},\theta)= 2\pi \hat{f}_0(\mathbf{x}).
\end{eqnarray}
Next, we represent the velocity at $\mathbf{x}$ by the complex number $v_0 e^{i\theta}$ whose real and imaginary part provide its $x$- and $y$- component, respectively.
Then, the averaged velocity field at $\mathbf{x}$ follows from the average of $v_0 e^{i\theta}$ with respect to the $N$-particle probability
\begin{eqnarray}
\frac{\langle\mathbf{v}(\mathbf{x})\rangle}{v_0}
&\equiv& \int d\mathbf{X}^{(N)} \int d\Theta^{(N)} P_N(\mathbf{X}^{(N)},\Theta^{(N)}) \int d\theta\, e^{i\theta}\Psi_1(\mathbf{x},\theta)
\\\nonumber
&=& N\int d\theta P_1(\mathbf{x},\theta) e^{i\theta}= 2\pi \hat{f}_1(\mathbf{x}).
\end{eqnarray}
We also consider spatial correlation functions for the densities
\begin{eqnarray}
\langle\rho(\mathbf{x}_1)\rho(\mathbf{x}_2)\rangle &\equiv&
\int d\mathbf{X}^{(N)} \int d\Theta^{(N)} P_N\left(\mathbf{X}^{(N)} \Theta^{(N)}\right)\int d\theta_1 d\theta_2\, 
\Psi_2\left(\mathbf{x}_1,\theta_1,\mathbf{x}_2,\theta_2\right)
\\\nonumber
&=&N(N-1)\int d\theta_1 d\theta_2 ~P_2(\mathbf{x}_1,\theta_1,\mathbf{x}_2,\theta_2)
\\\nonumber
&=& (2\pi)^2\Bigg[\left(1-\frac{1}{N}\right)\hat{f}_0(\mathbf{x}_1)\hat{f}_0(\mathbf{x}_2)+\hat{g}_{0,0}(\mathbf{x}_1,\mathbf{x}_2)\Bigg],
\end{eqnarray}
and for the velocities
\begin{eqnarray}
\frac{\langle\mathbf{v}(\mathbf{x}_1)\mathbf{v}(\mathbf{x}_2)\rangle}{v_0^2}&\equiv&
\int d\mathbf{X}^{(N)} \int d\Theta^{(N)}  P_N\left(\mathbf{X}^{(N)} \Theta^{(N)}\right)
\\\nonumber
&&~~~~~\int d\theta_1 d\theta_2 \mathrm{Re}\left(e^{i\theta_1}e^{-i\theta_2}\right)\Psi_2(\mathbf{x}_1,\theta_1,\mathbf{x}_2,\theta_2)
\\\nonumber
&=&N(N-1)\int d\theta_1 d\theta_2  \mathrm{Re}\left(e^{i\theta_1}e^{-i\theta_2}\right) P_2(\mathbf{x}_1,\theta_1,\mathbf{x}_2,\theta_2)
\\\nonumber
&=&(2\pi)^2\Bigg[\left(1-\frac{1}{N}\right)\frac{\hat{f}_1(\mathbf{x}_1)\hat{f}_{-1}(\mathbf{x}_2)+\hat{f}_{-1}(\mathbf{x}_1)\hat{f}_1(\mathbf{x}_2)}{2}
\\\nonumber
&&~~~~~~~~+\frac{\hat{g}_{1,-1}(\mathbf{x}_1,\mathbf{x}_2)+\hat{g}_{-1,1}(\mathbf{x}_1,\mathbf{x}_2)}{2}\Bigg].
\end{eqnarray}
Here, we used the representation of  
the dot product of two velocities by $\mathrm{Re}\left[\mathbf{v}(\mathbf{x})\mathbf{v}^*(\mathbf{y})\right]$ where $\mathbf{v}^*$ is complex conjugated to $\mathbf{v}$. 
For large $N$, one finds that the connected correlation function is simply, 
\begin{eqnarray}
\langle\rho(\mathbf{x}_1)\rho(\mathbf{x}_2)\rangle_\mathrm{c}
&\equiv& \langle\rho(\mathbf{x}_1)\rho(\mathbf{x}_2)\rangle
-\langle\rho(\mathbf{x}_1)\rangle\langle\rho(\mathbf{x}_2)\rangle
\\\nonumber
&=&4\pi^2\hat{g}_{0,0}(\mathbf{x}_1,\mathbf{x}_2)
\end{eqnarray}
for the densities and 
\begin{eqnarray}
\label{VELOC_CORR}
\langle\mathbf{v}(\mathbf{x}_1)\mathbf{v}(\mathbf{x}_2)\rangle_\mathrm{c}
&\equiv& \langle\mathbf{v}(\mathbf{x}_1)\mathbf{v}(\mathbf{x}_2)\rangle
-\langle\mathbf{v}(\mathbf{x}_1)\rangle\langle\mathbf{v}(\mathbf{x}_2)\rangle
\\\nonumber
&=&2\pi^2 v_0^2\big[\hat{g}_{1,-1}(\mathbf{x}_1,\mathbf{x}_2)+\hat{g}_{-1,1}(\mathbf{x}_1,\mathbf{x}_2)\big]
\end{eqnarray}
for the  velocities.
For homogeneous states where translational invariance applies, one defines the integrated correlation function 
$\int_\mathrm{all} d\mathbf{x}~\langle \phi(\mathbf{x})\phi(\mathbf{x}+\mathbf{r})\rangle$.
One can also calculate the "specific" correlation function, the correlation normalized by the number of ordered pairs, by dividing the correlation by $N(N-1)$.
This will allows us to closely compare systems composed of different number of particles. 
In experiments, it is quite often that the velocity correlation function measured is usually not weighted by the density correlation as defined here. To achieve the non-weighted velocity correlation, we divide the velocity correlation by the density correlation.

Next, we consider global quantities. 
We define a complex order parameter $\Omega$ for a single realization of the system at a given time,
\begin{equation}
\label{ORDER_P_DEF}
\Omega \equiv \frac{1}{N}\sum_{j=1}^N e^{i\theta_j}.
\end{equation}
where we sum up the normalized complex velocities of all particles.
The ensemble average of $\Omega$ follows as 
\begin{eqnarray}
\label{OMEGA_AV}
\langle\Omega\rangle&=&\int d\mathbf{x}^{(N)} d\theta^{(N)} \Omega ~P_N\left(\mathbf{x}^{(N)},\theta^{(N)}\right)
\\\nonumber
&=&\frac{2\pi}{N}\int d\mathbf{x}\hat{f}_1(\mathbf{x})
\end{eqnarray}
The norm of the order parameter squared is 
\begin{eqnarray}
|\Omega|^2 &=& \Omega\Omega^*
\\\nonumber
&=&\frac{1}{N}+\frac{1}{N^2}\sum_{j\neq k}e^{i(\theta_j-\theta_k)},
\end{eqnarray}
and its ensemble average
\begin{equation}
\label{OMEGA_VAR}
\langle|\Omega|^2\rangle=\frac{1}{N}+\frac{N-1}{N}\langle\Omega\rangle\langle\Omega^*\rangle+\frac{(2\pi)^2}{N^2}\int d\mathbf{x}_1 d\mathbf{x}_2~
\frac{\hat{g}_{1,-1}(\mathbf{x}_1,\mathbf{x}_2)+\hat{g}_{-1,1}(\mathbf{x}_1,\mathbf{x}_2)}{2}.
\end{equation}
The second term comes from the average taken with respect to the first term in the Ursell-expansion, 
$P_1(z_1)\cdots P_1(z_j)\cdots P_1(z_k)\cdots P_1(z_N)$, while the last term comes from $P_1(z_1)\cdots G_2(z_j,z_k)\cdots P_1(z_N)$. 
For large $N$, the variance of the order parameter becomes 
\begin{equation}
\label{VAR_DEF}
\langle|\Omega-\langle\Omega\rangle|^2\rangle=\frac{(2\pi)^2}{N^2}\int d\mathbf{x}_1 d\mathbf{x}_2~
\frac{\hat{g}_{1,-1}(\mathbf{x}_1,\mathbf{x}_2)+\hat{g}_{-1,1}(\mathbf{x}_1,\mathbf{x}_2)}{2}.
\end{equation}
From Eq. (\ref{OMEGA_AV}), we see that the averaged order parameter is related to the 
first mode of the one-particle density distribution. 
It is zero if the total momentum vanishes and reflects nothing about local orientational or positional order. 
The variance contains information about pairwise correlations. 
The lowest order of the local organization is revealed by this quantity, 
which is not necessarily zero in the disordered state. 
According to Eq. (\ref{VELOC_CORR}), the equal-time connected velocity correlation function is given by 
the Fourier coefficients $\hat{g}_{\pm 1,\mp 1}$. 
Thus, the variance of the order parameter, Eq. (\ref{VAR_DEF}), can be interpreted as the spatial integral 
over the connected velocity correlations.
\subsection{Conservation laws}
\label{sec.conservation_laws}
We have seen in the previous section that $\hat{f}_0(\mathbf{x}_1)$ is given by  
the local density at $\mathbf{x}_1$, and that $\hat{g}_{00}(\mathbf{x}_1,\mathbf{x}_2)$ is proportional to the connected density correlation at $\mathbf{x}_1$ and $\mathbf{x}_2$. These two quantities should be conserved by the collision operator. 
This is because instantaneous collisions only change velocities but not the positions of particles. 
Thus, densities and their correlations can only change in the streaming step.
We now inspect the conservation laws regarding these two quantities. 
The coupling constants, Eqs.(\ref{eq.coupling_definition}), have the general form,
\begin{equation}
w(m_1, m_2,\cdots, p_1, p_2,\cdots)=\prod_{k}\int \frac{d \theta_k}{2\pi}e^{ip_k \theta_k}\prod_{j}e^{-im_j \Phi_j},
\end{equation}
where $m_j$ is the mode number with respect to the post-collision angle whereas $p_j$ refers to the pre-collision angle. 
When all the $m_j$'s are zero, all pre-collisional mode numbers must also vanish,
\begin{equation}
w(m_1=0, m_2=0\cdots, p_1, p_2\cdots)=\prod_{k}\delta_{p_k,0}.
\end{equation}
Hence, one has a relatively simple post-collision formula, where for both $\hat{f}'_0(\mathbf{x}_1)$ and $\hat{g}'_{0,0}(\mathbf{x}_1,\mathbf{x}_2)$, 
only the zero modes
$\hat{f}_0(\mathbf{x}_1)$ and $\hat{g}_{00}(\mathbf{x}_1,\mathbf{x}_2)$ contribute.
According to equation (\ref{eq.marginalization_condition}), one also has
\begin{equation}
\int_\mathrm{all} d\mathbf{x}_2 ~\hat{g}_{00}(\mathbf{x}_1,\mathbf{x}_2)=0.
\end{equation}
This condition eliminates 
all those terms in the series expansion, Eqs. (\ref{eq.1st_BBGKY_full}, \ref{eq.2nd_BBGKY_full}), 
that involve at least one spatial integration of $\hat{g}_{00}(\mathbf{x}_1,\mathbf{x}_2)$. For $N\rightarrow \infty$, one eventually arrives at
\begin{equation}
\hat{f}'_0(\mathbf{x}_1)=e^{-M(\mathbf{x}_1)}\sum_{p=0}^{\infty} \frac{N!}{p!(N-p)!}\left(\frac{M}{N}\right)^p \hat{f}_0(\mathbf{x}_1)=\hat{f}_0(\mathbf{x}_1),
\label{eq.f0_conservation}
\end{equation}
because $N!/(N-p)!\rightarrow N^p$ for $N\rightarrow \infty$ and $\sum_p^{\infty} M^p/p!={\rm e}^M$.
Similarly, one finds
\begin{equation}
\hat{g}'_{00}(\mathbf{x}_1,\mathbf{x}_2)=\hat{g}_{00}(\mathbf{x}_1,\mathbf{x}_2).
\end{equation}
This means that if we were to sum diagrams to infinite order, the conservation laws would be fulfilled.
However, our low density expansions, Eqs. (\ref{INFINITE_N_FIRST}, \ref{INFINITE_N_SECOND}) include only a limited number of diagrams and expand the exponential prefactors.
It turns out that even these truncated expressions do not violate 
the conservation laws as long as the expansion is consistent, that is, all terms up to a given order $S$ in $M^S$ are included.
In this case, terms that would violate the conservation laws cancel each other exactly at each order in $M$.
Therefore, the conservation laws provide a consistency test of the low density expansions.

Now let us inspect the conservation law for finite $N$ for the first hierarchy equation. 
The generalization to the second equation can be done by a similar approach.
For finite $N$, equation (\ref{eq.f0_conservation}) turns into 
\begin{equation}
\hat{f}'_0(\mathbf{x}_1)=\sum_{p=0}^N \frac{N!}{p!\,(N-p)!}\left(\frac{M}{N}\right)^p \left(1-\frac{M}{N}\right)^{N-p}\hat{f}_0(\mathbf{x}_1)
\end{equation}
Because of the binomial formula,
\begin{equation}
\label{BINOMIAL_F}
1=1^N=\left(1-\frac{M}{N}+\frac{M}{N}\right)^N=\sum_{p=0}^N \frac{N!}{p!\,(N-p)!}\left(\frac{M}{N}\right)^p \left(1-\frac{M}{N}\right)^{N-p}
\end{equation}
the conservation law is fulfilled, $\hat{f}'_0(\mathbf{x}_1)=\hat{f}'_0(\mathbf{x}_1)$.
Similar to the case of infinite $N$, 
it is easy to see that the conservation laws remain fulfilled if one truncates the BBGKY equations
in a consistent way \cite{FOOT3}, that is by including all terms up to given order $O(M/N)^S$ and neglecting the rest. 

\section{Numerics}
\label{sec:numerics}

\subsection{Algorithm}
\label{sec:algorithm}

In this section we outline the numerical solution of the BBGKY-hierarchy equations. 
Analytical solutions will be left for future work.
Here, we focus on spatially homogenous solutions. 
For homogeneous states, the coefficients $\hat{f}_p$ are independent of position
and the coefficients for the two-particle correlations depend only on the difference of the spatial arguments,
\begin{equation}
\hat{g}_{mn}({\bf x}_1, {\bf x}_2)\equiv \hat{g}_{mn}({\bf z})\,,\;\;\;{\rm with}\;{\bf z}={\bf x}_2-{\bf x}_1\,. 
\end{equation}
This reduces the dimensionality of the space for $\hat{g}_{mn}$ from four to two.
We also assume isotropic states, where $\hat{f}_0=\rho_0/(2\pi)$ and $\hat{f}_k=0$ for $k\ge 1$ \cite{FOOT4}.
This solves the first BBGKY-equation exactly, and we only have to deal with the second hierarchy equation.

Using the reduced space variable ${\bf z}={\bf x}_2-{\bf x}_1$, the second BBGKY equation can be written symbolically as
\begin{equation}
\label{NUM_STREAM1}
g_2({\bf z},\theta_1,\theta_2,t+\tau)=C({\bf z}',\theta_1,\theta_2,t)
\end{equation}
where $C$ denotes the collision integral evaluated at the ``back-streamed'' 
position ${\bf z}'={\bf z}-\tau({\bf v}_2(\theta_2)-{\bf v}_1(\theta_1))$. 
We solve this equation numerically by a method that 
is related to the one from Ref. \cite{ihle_13}. 
The main idea is to explicitly perform the streaming step for the function $g_2$ on a cubic grid 
while the collision operator is 
evaluated in angular Fourier space,
\begin{equation}
\label{NUM_COLL1}
C({\bf z}',\theta_1,\theta_2)=\sum_{m,n}\hat{C}_{mn}
(\mathbf{z}')~e^{i m \theta_1}e^{i n\theta_2}~~.
\end{equation}
The Fourier coefficients $\hat{C}_{mn}$ 
follow from the Fourier transformation of the diagrammatic equation, Eq. (\ref{INFINITE_N_SECOND}).
Thus, $\hat{C}_{mn}$ is a composed of diagrams such as, for example, $\big\langle\sDaGjf\big\rangle_{mn}$ for the 
strong-overlap case with $|{\bf z}'|\le R$, or  
$\big\langle\sDbfGk\big\rangle_{mn}$ for $R< |{\bf z}'|\le 2R$, 
or $\big\langle\sDcfGi\big\rangle_{mn}$ for the no overlap region, $|{\bf z}'|> 2R$.

The reduced space variable ${\bf z}$ is discretized on a grid with $L\times L$ points and periodic boundary conditions. Its x- and y-coordinates run from
$-L/2$ to $L/2$ respectively. 
Typical values for $L$ were between 36 and 100 lattice units. 
In our algorithm, the Fourier modes $\hat{g}_{p,q}$
of the connected two-particle correlation function are stored at every point of the grid.
We mostly used Fourier series up $\pm 11$ modes, 
i.e. we include all modes with $-11\le m,~n\le11$, but 
in few cases with very small noise, $\pm 21$ modes were used.
The results are tested to be converged to the series where higher modes were included.
At the beginning of each iteration, at every grid point the corresponding diagrams from Appendix B
are calculated. For example, for all grid points ${\bf z}$ that are closer to the origin than the radius $R$, the diagrams with strong-overlap are needed. 

To evaluate the diagrams, the quantities $\bar{G}_{mn}$, $\Delta\bar{G}_{mn}$, and $\bar{\bar{G}}_{mn}$, see Eqs. (\ref{FIRST_BAR}, 
\ref{SECOND_BAR}, \ref{THIRD_BAR}) must be calculated. This requires spatial integrations of $\hat{g}_{mn}$ 
over circles and intersections of circles.
The integrals are 
found
by interpreting them as spatial averages over these domains.
For example, according to Eq. (\ref{FIRST_BAR}), and since $\hat{g}_{mn}(\mathbf{x}_1,\mathbf{x}'_2)$ is equivalent to $\hat{g}_{mn}(\mathbf{x}'_2-\mathbf{x}_1)$, we obtain $\Delta\bar{G}_{mn}(\mathbf{z})$ by integrating over a 
circle which is centered around the reduced location ${\bf z}=(x,y)$: 
\begin{equation}
\label{FIRST_BAR_INTEG}
\Delta\bar{G}_{mn}(\mathbf{z})=\int_{\odot} \hat{g}_{mn}(\mathbf{z}')\,d \mathbf{z}'
\approx {\pi R^2 \over N_1}\,\, \sum_{(i-x)^2+(j-y)^2\le R^2} \hat{g}_{mn,ij}
\end{equation}
Here, the integral is evaluated by summing up values from all grid points $(i,j)$ that are inside a circle of radius $R$. 
This sum is divided
by the number $N_1$ of these grid points
and multiplied with the area $\pi R^2$ of the domain.
To ensure accurate integration, $R$ must be large enough. 
We used values of $R$ ranging from 3 to 24 lattice units.
Once all diagrams have been determined, the coefficients $\hat{C}_{mn}$ are calculated.
The goal of an iteration step is to determine the new coefficients $\hat{g}_{m,n}$. 
To do this, we first obtain $g_2$ in real space, that is $g_2({\bf z},\theta_1,\theta_2)$. 
Both angles $\theta_1$ and $\theta_2$ are discretized into $P=64$ equidistant points on the interval $[0,2\pi]$.
For a given grid point ${\bf z}$ and for every value of the allowed angles, we ``back-stream'' to the point 
${\bf z}'={\bf z}-\tau({\bf v}_2(\theta_2)-{\bf v}_1(\theta_1))$. 
At this off-lattice point, we obtain the coefficients
$\hat{C}_{mn}$ by interpolation from the known values at adjacent grid points. Using Eq. (\ref{NUM_COLL1}), we  
reconstruct the real space value of the
collision operator and, following Eq. (\ref{NUM_STREAM1}), we equate this with $g_2({\bf z},\theta_1,\theta_2,t+\tau)$. 
Once this is done for all permitted 
back-stream 
vectors for a particular location ${\bf z}$, the updated coefficents $\hat{g}_{m,n}({\bf z},t+\tau)$ are 
extracted via angular Fourier transformation.
Note, that this procedure involves an angular filtering, because we implicitly set higher Fourier modes with 
$|m|,~ |n|>11$ to zero.

The algorithm can be accelerated by using the assumed homogeneity and isotropy of the system.
In this case,
one can show that only the coefficients $\hat{g}_{n,-n}$ are non-zero, which significantly reduces 
the number of modes to be updated.
To eliminate the build-up of eventual discretization errors, after every iteration we explicitly enforce the normalization condition, 
Eq. (\ref{eq.marginalization_condition}).
In terms of Fourier-modes, this amounts to applying tiny homogeneous shifts to the coefficients
$\hat{g}_{00}$, $\hat{g}_{0n}$ and $\hat{g}_{n0}$, $n=1,2,\ldots$,
such that their integrals over the entire simulation box vanish.

Initializing the system with an uncorrelated, ideal gas-like state where all $\hat{g}_{m,n}$ vanish,
one first observes the build-up of correlations inside the collision zone $|{\bf z}|\le R$. 
These correlations are then spreading outside
the zone due to streaming, and correlated collisions will continue to happen until a stationary state is reached.
Applications of this algorithm will be presented in Section ~\ref{sec:results}.

\subsection{Measurements and verification}
\label{sec:verify}

To verify the numerical approach and to test the general validity of the ring-kinetic formalism,
we perform detailed comparisons with agent-based simulations.
To enable meaningful comparisons, one has to identify appropriate parameter ranges and highly diagnostic observables.
For example, the order parameter $\Omega$ and its variance were defined in Section ~\ref{sec.physical_quantities} 
such that,
on one hand, they have a simple relation to the lowest Fourier coefficients of the kinetic theory and, on the other hand, can be easily measured in agent-based simulations. 

Natural systems of self-propelled particles such as swarms of fish, bird, insects or 
bacteria have small particle numbers of order $10^1$ to
$10^4$. 
For example, the wild swarms of midges, recently investigated by Attanasi {\em et al.} \cite{attanasi_14}, only contain 100 to 600 midges. Studying
swarms of {\em Chironomus riparius} midges, Puckett and Ouellete \cite{puckett_14} even found
that once the swarms contain order 10 individuals, 
all statistics saturate and the swarms enter an asymptotic regime.

Thus, the idea of the thermodynamic limit $N\rightarrow \infty$ which, in regular statistical mechanics, 
is motivated by the large number of atoms, $>10^{23}$, in condensed matter systems, is not always useful here. 
Therefore, investigating the effects of small particle numbers in active matter systems is worthwhile.
Furthermore, practical limitations of the kinetic theory algorithm also force us to run agent-based simulations at 
small particle numbers, $2\le N\le 100$, and to put more emphasis on the variance of the order parameter $\Omega$ 
(defined in Eq. (\ref{ORDER_P_DEF})) instead of $\langle \Omega \rangle$.
This is because,  
on one hand, in the numerical algorithm for the BBGKY-equations, 
the radius $R$ must be well discretized  by a sufficiently large number of grid points. On the other hand,
the ratio $L/R$ must also be sufficiently large in order to minimize artifacts to the periodic boundaries and to enable 
the observation of possible power law decay of the correlations. We use $L/R=3\ldots 33$. This  fixes the choice of the linear system size 
$L$. 
However, choosing $L$ too large will be computationally unfeasible. As a compromise we arrive at maximum lengths around $L=100$. 

Another restriction is imposed by the low density expansion which requires that the average partner number $M$ should be small.
Given that the restrictions are coupled via $M=\pi R^2\rho=\pi N (R/L)^2$, we find that the total particle number must be quite small, $N\le 50$, to ensure sufficient accuracy at realistic computational times on an eight core CPU.
Therefore, agent-based simulations with small particle numbers must be performed to allow for direct comparison. 
When $N$ is small, even if there is strong global order and all particles are more or less aligned, 
the direction of the total momentum vector will
rather rapidly fluctuate in the agent-based simulations. 
Time- or ensemble averaging $\Omega$ will eventually lead to $\langle \Omega \rangle= 0$ and hence 
$\hat{f}_1= 0$. 

This is different from 
the thermodynamic limit $N\rightarrow \infty$ and agent-based simulations at very large particle numbers.
At large $N$, the direction of collective motion is usually pinned by the underlying square simulation box and goes into the $(\pm 1,0)$, 
$(0,\pm 1)$ or $(\pm 1, \mp 1)$ directions. The probability for the global direction to switch within the simulation time is small, and time-averages in the ordered phase will lead to nonzero $\langle \Omega \rangle$ and $\hat{f}_1\neq 0$.
Hence, in our case of small $N$, 
we use the variance of $\Omega$ to describe global order. If $\langle \Omega \rangle=0$, the variance becomes
$var(\Omega)=\langle |\Omega|^2\rangle$ which remains an informative quantity down to $N=2$.

To obtain more detailed insight than a global quantity like $\Omega$ can deliver, we also measure the following correlation functions according to the definitions of Section \ref{sec.physical_quantities}.
First, we define the connected integrated correlation function per ordered pair for the density
\begin{eqnarray}
\label{eq.C_d}
C_{\rho}(\mathbf{r})&\equiv&\frac{1}{N(N-1)}\int_\mathrm{all} d\mathbf{x}~\langle \rho(\mathbf{x})\rho(\mathbf{x}+\mathbf{r})\rangle_c
\\\nonumber
&=& \frac{4\pi^2}{N(N-1)} \int_\mathrm{all}d\mathbf{x}\hat{g}_{0,0}(\mathbf{x},\mathbf{x}+\mathbf{r})~,
\end{eqnarray}
and for unit velocity
\begin{eqnarray}
\label{eq.C_v}
C_v(\mathbf{r})&\equiv&\frac{1}{N(N-1)v^2}\int_\mathrm{all} d\mathbf{x}~\langle \mathbf{v}(\mathbf{x})\mathbf{v}(\mathbf{x}+\mathbf{r})\rangle_c
\\\nonumber
&=& \frac{2\pi^2}{N(N-1)} \int_\mathrm{all}d\mathbf{x}\big[\hat{g}_{1,-1}(\mathbf{x},\mathbf{x}+\mathbf{r})+\hat{g}_{-1,1}(\mathbf{x},\mathbf{x}+\mathbf{r})\big]~.
\end{eqnarray}
We also define the non-weighted connected integrated correlation function for unit velocity
\begin{eqnarray}
\label{eq.G_v}
G_v(\mathbf{r})&=&
\frac{\int_\mathrm{all} d\mathbf{x}~\langle \mathbf{v}(\mathbf{x})\mathbf{v}(\mathbf{x}+\mathbf{r})\rangle_c}
{\int_\mathrm{all} d\mathbf{x}~\langle \rho(\mathbf{x})\rho(\mathbf{x}+\mathbf{r})\rangle}
\\\nonumber
&=&\frac{ \int_\mathrm{all}d\mathbf{x}\big[\hat{g}_{1,-1}(\mathbf{x},\mathbf{x}+\mathbf{r})+\hat{g}_{-1,1}(\mathbf{x},\mathbf{x}+\mathbf{r})\big]}
{2v^2\int_\mathrm{all}d\mathbf{x}\left[\left(1-\frac{1}{N}\right)\hat{f}_{0}(\mathbf{x})\hat{f}_0(\mathbf{x}+\mathbf{r})+\hat{g}_{0,0}(\mathbf{x},\mathbf{x}+\mathbf{r})\right]}
~.
\end{eqnarray}

\section{Results}
\label{sec:results}

In this section we give numerical results for the ring-kinetic theory and compare with agent-based simulations.
We begin by studying a $2$-particle system
because the theory is supposedly exact for $N=2$. 
Using the collision terms given diagrammatically by Eq. (\ref{FINITE_N_SECOND}) we follow the algorithm outlined
in section {\ref{sec:algorithm}:
Eq. (\ref{NUM_STREAM1}) for the two-particle correlation function $g_2$ is iterated numerically
until a stationary state is reached.
The lowest Fourier-modes $\hat{g}_{0,0}$, $\hat{g}_{1,-1}$ and $\hat{g}_{-1,1}$ are extracted from $g_2$ by means of 
Eq. (\ref{FOURIER_LABEL}) and then used to calculate the integrated correlation functions $C_{\rho}$, $C_v$ and $G_v$ 
according to Eqs. (\ref{eq.C_d}--\ref{eq.G_v}).
In addition, agent-based simulations of Eqs. (\ref{VM_UPDATE}--\ref{ANGLE_RULE})
in a square box with periodic boundary conditions  
were also performed. 

Measurements of the correlation functions were taken {\em after} the streaming step, e.g. 
in the pre-collisional state, in order to match the kinetic theory predictions.
For zero particle velocity in Figs. \ref{fig.Cd-N2} and \ref{fig.Cv-N2}, these measurements were ensemble-averaged over
$4\times 10^9$ realizations whereas for nonzero speed averages over $10^{11}$ realizations were performed.
The error bars in Figs. \ref{fig.Cd-N2}--\ref{fig.Gv-N2} are smaller than the size of the symbols.
As shown by these figures,
the predictions of kinetic theory are in perfect quantitative agreement with agent-based simulations.
The results of the connected density correlation function for various mean-free path are shown in Fig. \ref{fig.Cd-N2}.
Comparing cases where $v>0$ with the case of vanishing speed, $v=0$, 
we see that streaming induces clustering: the particles develop a tendency to stay closer to each other 
than in an uncorrelated gas. 
This effect shows as a positive density correlation $C_{\rho}$ inside the collision circle ($r\le R$) and  $C_{\rho}<0$ outside ($r>R$). 
Note, that negative density correlations are necessary to compensate for the positive ones, 
since the integral of $\hat{g}_{00}$ over the entire 
volume must be zero to fulfill the normalization requirement, Eq. (\ref{eq.marginalization_condition}).

One also observes that the smaller the speed is, the larger is the correlation inside the collision circle. 
This is because the particles with larger speeds have a large chance to escape from each other and 
hence clusters are more likely to break apart.
This also implies that the case of very small speed is qualitatively different from zero speed.
At small speeds, correlations and clusters will build up very slowly but finally become large in the steady state, whereas 
clusters can never form when particles are not permitted to move at all.
Fig. \ref{fig.Cd-N2} is thus consistent with the conjecture expressed by many researchers, see for example \cite{czirok_97,baglietto_09b},
that the $v_0=0$ case is a singular limit: there seems to be no smooth transition from the equilibrium Heisenberg-like model 
at $v_0=0$ to the noneqilibrium VM at $v_0>0$.
Note, that even though the normalized density $M$ is not small in Figs. \ref{fig.Cd-N2}--\ref{fig.Gv-N2}, 
agreement is still perfect. This is because no density expansion is necessary for $N=2$, all diagrams are included
and the higher n-particle correlations such as $G_3$ are naturally zero.
\begin{figure}
\begin{center}
\includegraphics[width=3.1in,angle=-90]{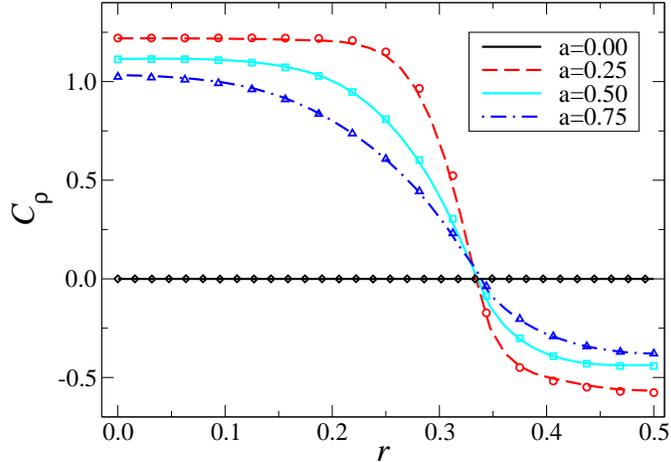}
\caption
{
The connected density correlation function defined in eq. (\ref{eq.C_d}) for systems with two particles.
The kinetic theory results are obtained by an iteration algorithm for the two-particle correlation function $g_2$
as described in section \ref{sec:algorithm}. 
The lines show the results of numerical evaluation of the theory and the open symbols the agent-based simulations. 
The system's linear size is 72 lattice units and rescaled to $L=1$ in the plot. 
Other parameters are $M=2\pi/9=0.6981$, $\eta=1.0$, $R/L=1/3$. The ratios of the mean-free path to the collisional 
radius $a=\tau v_0/R$ are given in the legend. 
For the $v_0=0$ case, in the agent-based simulations 
the initial locations of the particles are randomly choosen
and an ensemble average over the different initializations is performed. 
In Section ~\ref{sec.conservation_laws} we argue that $\hat{g}_{00}(z'_1,z'_2)=\hat{g}_{00}(z_1,z_2)$. 
We initialized $\hat{g}_{00}(z_1,z_2)$ to be zero, which corresponds to a Poissonian particle distribution.
In both simulation and theory, $\hat{g}_{00}(z_1,z_2)$ 
cannot adjust at exactly zero speed.  
}
\label{fig.Cd-N2}
\end{center}
\end{figure}
For the connected velocity correlation function (see Fig. \ref{fig.Cv-N2}), 
we see that streaming ``switches on'' correlations outside the collision circle. 
This means the information has been spread out. 
The larger the speed, the further the information is spread and the stronger the correlations can be built up outside the collision zone. 
The payoff is that the correlation within the interaction range $R$ is reduced for large speed. 
That means, subsequent collisions (that only take place among particles within interaction range) 
will be less correlated. This is consistent with our hypothesis that large ratios $\tau v_0/R$ 
will make the behavior more mean-field-like. 

Comparing results with and without streaming, we find that streaming dramatically increases the velocity correlations 
inside the circle. 
We suspect that this is again caused by clustering which increases the probability of 
finding one particle inside the collision circle of the others. Therefore, particles ``see'' a local environment corresponding to
a system of higher density. This means, even at $M\ll 1$, most particles have several potential partners they 
travel and 
repeatedly collide with
instead of only occasionally capturing a partner which would lead to a quick decorrelation of velocities after 
the particles have left interaction range. 
\begin{figure}
\begin{center}
\includegraphics[width=3.1in,angle=-90]{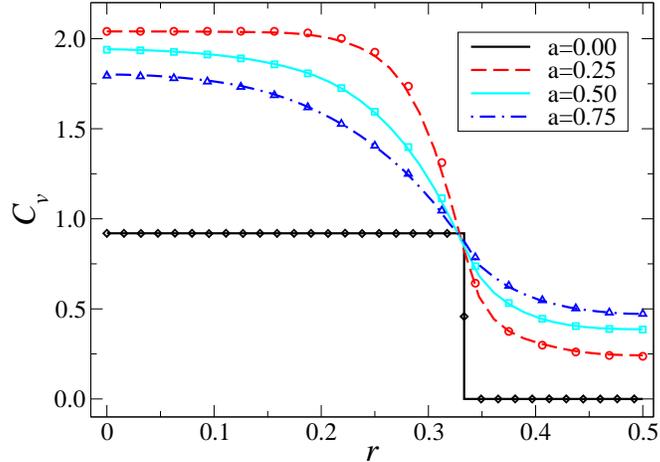}
\caption
{
The connected velocity correlation function, Eq. (\ref{eq.C_v}), for $2$-particle systems.
The parameters are the same as Fig. \ref{fig.Cd-N2}.
}
\label{fig.Cv-N2}
\end{center}
\end{figure}
To directly calculate the velocity correlation of two particles without taking into account the possibility of 
finding them in specified locations, we look at the non-weighted correlation function Fig. \ref{fig.Gv-N2}. 
This plot clearly shows that the non-weighted velocity correlation 
cannot be larger than the one of immobile but interacting agents 
(the black dashed line and symbols for $r\le R$). The decrease of $G_v$ 
inside the collision zone is a result of the influx of  particles from outside the interaction range, 
as seen from the point of view of the focal particle. 
The most efficient way of decreasing the correlation is through the head-on collision of two particles. 
This means that the correlation in a region which extends a distance $2\tau v=2aR$ from the circumference inward, 
will be reduced when streaming is turned on. By inspecting carefully the inset of Fig. \ref{fig.Gv-N2}, 
we see that our results quantitatively confirm this reduction effect. 
The red curve which 
corresponds to $2\tau v=R/2$ starts to decrease below the $v=0$ curve at $r=R/2$, and the cyan curve, where $2\tau v=R$, 
starts decreasing already at $r=0$.
\begin{figure}
\begin{center}
\includegraphics[width=3.1in,angle=-90]{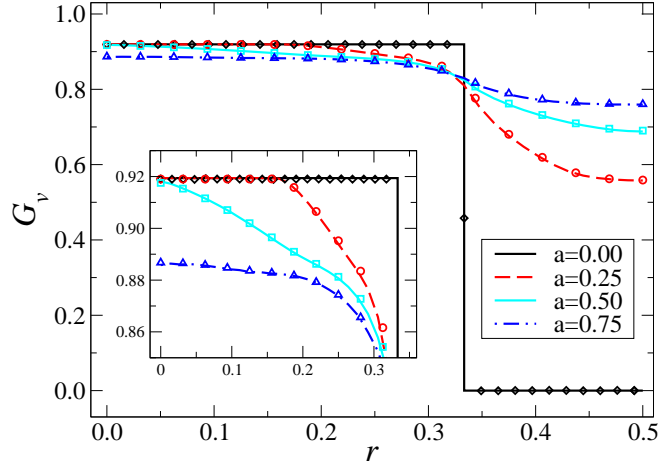}
\caption
{
The non-weighted connected velocity correlation function, eq. (\ref{eq.G_v}), for $2$-particle systems.  
The system's parameters are the same as Fig. \ref{fig.Cd-N2}. The inset shows the correlation for short distance.
}
\label{fig.Gv-N2}
\end{center}
\end{figure}

Next, we look at the results for a $5$-particle systems with a 
relatively small $R/L$ ratio in Fig. \ref{fig.Gv-N5}. 
Again, the theory excellently agrees with the simulations, although due to the low-density expansion, diagrams with four 
and five particles are neglected.
Remember that the multi-particle correlations $G_3$, $G_4$ and $G_5$, which do exist in a $N=5$ system, 
are also neglected in our theory.
Therefore, Fig. \ref{fig.Gv-N5} is the first indication that the ring-kinetic theory for Vicsek-like models 
can deliver quantitatively correct results,
at least in not too strongly-correlated regimes. 
For large $\tau v/R$, the long-distance correlations show small oscillations (see the red and cyan curves) which are well
reproduced by kinetic theory. 
This effect is usually observed when $\tau v \ge R$. The oscillation becomes more apparent as the noise is increased although the over-all correlation is reduced. 
For small $\tau v/R$ (blue), there is a maximum correlation near the boundary of the collision circle. 
We hypothesize that both oscillations and the maximum could be
resonance effects caused by the 
fixed distance $\lambda$, particles travel in each time step.

We also compare the velocity correlation function for systems with different number of particles but with the 
same $R/L$ and $\tau v/R$ ratio (Fig. \ref{fig.Gv-N10-20}). The long-distance behavior for the velocity correlation function are found to collapse into a master curve.
\begin{figure}
\begin{center}
\includegraphics[width=3.1in,angle=0]{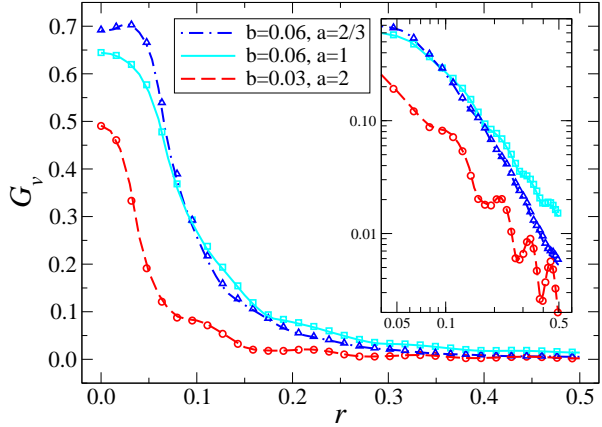}
\caption
{
The non-weighted connected velocity correlation function, eq. (\ref{eq.G_v}), for systems with $N=5$.
The lines show the results of numerical evaluation of the theory and the symbols the agent-based simulations.
The system's noise is $\eta=1.5$. 
The system size is 100 lattice units for the cyan and red curve, 150 lattice units for the blue curve but was rescaled to $L=1$ in the plot.
The ratio of collisional radius to the system's linear size $b=R/L$, and mean-free path to the radius $a=\tau v/R$ 
are indicated by the legend. 
For $b=0.06$, $M=\pi N b^2$ is equal to $0.0565$, while for the run with $b=0.03$ we have $M=0.0141$.
The inset shows the same data but in $\log$-$\log$ scale. 
}
\label{fig.Gv-N5}
\end{center}
\end{figure}
\begin{figure}
\begin{center}
\includegraphics[width=3.1in,angle=-90]{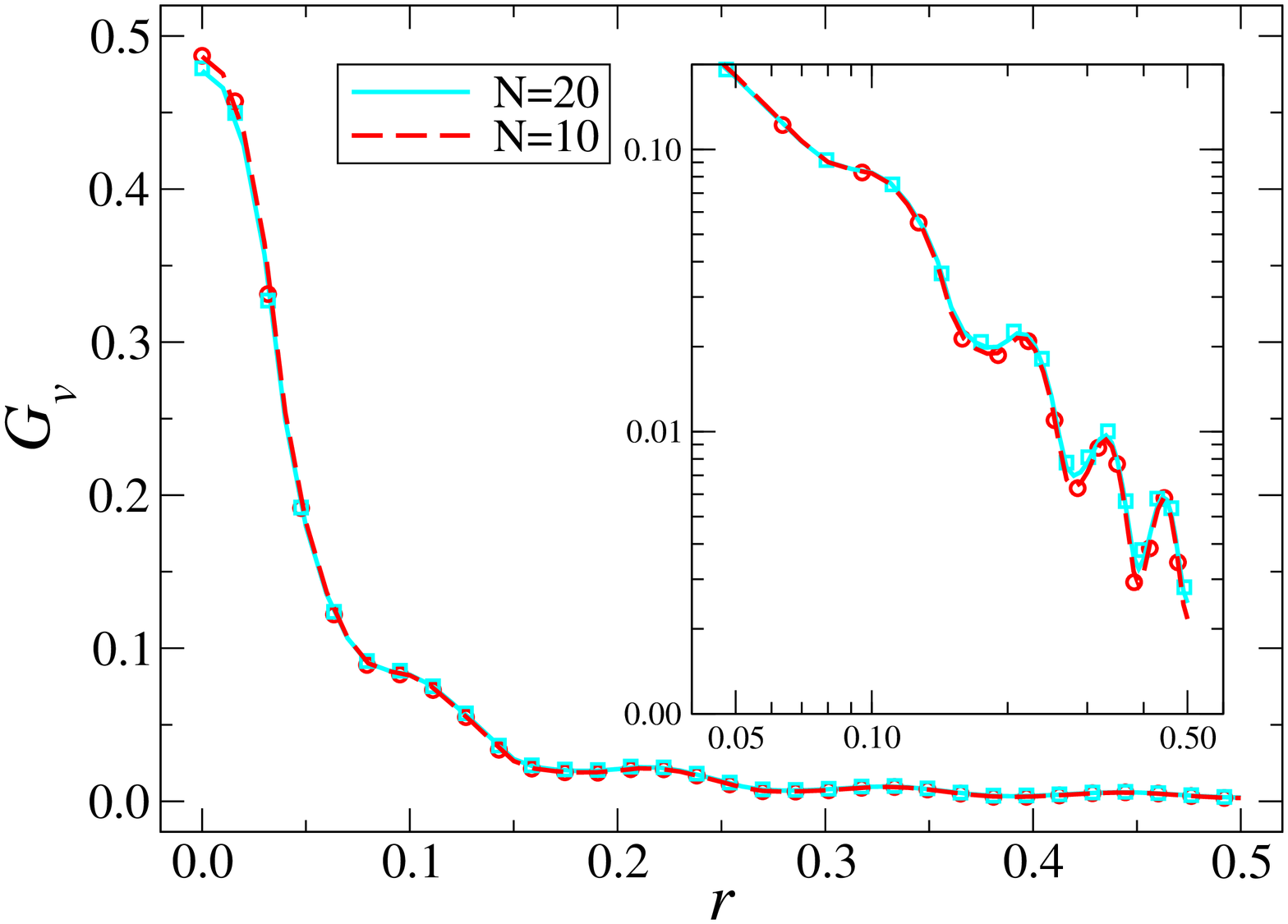}
\caption
{
The non-weighted connected velocity correlation function, eq. (\ref{eq.G_v}), for $10$-particle, and $20$-particle systems.  
The system's parameters are  are $\eta=1.5$, $R/L=0.03$, and $\tau v/R=2$. The inset shows the same data but in $\log$-$\log$ scale.
The system size is fixed to 100 lattice units but rescaled to $L=1$ in the plot.
}
\label{fig.Gv-N10-20}
\end{center}
\end{figure}
As observed in Fig. \ref{fig.Cv-N2} (and also Fig. \ref{fig.Gv-N2}), where the velocity correlation function decreases inside but increases outside the collision circle as the speed of the particle increases, there might be a optimized $\tau v/R$ ratio where the correlation can be spread most effectively across the system. 
To have a better understanding regarding this aspect we studied more global aspects of velocity correlation. 
As discussed in section \ref{sec.physical_quantities}, the integrated velocity correlation function is 
proportional to the variance of the order parameter. We define two related quantities here:
The connected velocity correlations integrated over all space
\begin{equation}
\mu=\frac{(2\pi)^2}{N^2}\int_{\mathrm{all}} d\mathbf{x}_1 \int_{\mathrm{all}}d\mathbf{x}_2~\hat{g}_{1,-1}(\mathbf{x}_1,\mathbf{x}_2),
\\\nonumber
\end{equation}
and integrated only over the collision circle,
\begin{equation}
\mu_c=\frac{(2\pi)^2}{N^2}\int_{\mathrm{all}} d\mathbf{x}_1 \int_{O_1}d\mathbf{x}_2~\hat{g}_{1,-1}(\mathbf{x}_1,\mathbf{x}_2).
\\\nonumber
\end{equation}
The results for agent-based simulations for $N=5$, $\eta=1.5$ and $M=0.0565$ are shown in Fig. \ref{fig.var_Omega}. 
Eq. (\ref{BVM_CRIT}) gives the mean-field prediction for the critical noise , $\eta_C(M=0.0565)\approx 0.61$, 
which is an upper bound of the actual
critical noise. Since we have $\eta=1.5>\eta_C$ we know
that the system investigated here corresponds to the disordered state \cite{FOOT5}. 
Nevertheless, the variance of the order parameter indicates that there is still some degree of 
local ordering. The maximum $\mu$ is found for systems with $\tau v=R$. 
For systems with $\tau v>R$, although the system strongly spreads the correlation to the outside of the collision circle, 
the variance decays with increasing $\tau v_0/R$. 
This is because the source where correlations are generated -- the collision zone --  was also burlily disturbed 
by incoming particles and by the departure of previous collision partners. 
However, decreasing the ratio $\tau v_0/R$ to below unity, 
reduces
the variance due to the inability to effectively transport correlations to the outside
of the collision zone.
We next look at the variance $\mu_c$ which is calculated with respect to the collision circle.
The data indicates that $\mu_c$ seems to decay exponentially for $\tau v<R$. However, for $\tau v>R$ 
there is a sudden qualitative change: 
the decay of $\mu_c$ becomes consistent with a power-law.
\begin{figure}
\begin{center}
\includegraphics[width=2.8in,angle=-90]{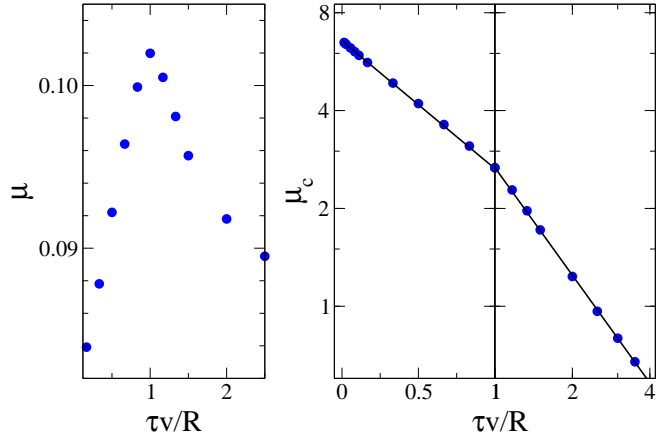}
\caption
{
Agent-based simulations for $5$-particle systems showing
the connected velocity correlations integrated over all space, $\mu$, (left panel) and integrated 
over the collision zone $\mu_c$ (right panel).
Note that on the right panel, the $x$-axis change from normal scale to log scale at $\tau v=R$ while the $y$-axis is in log scale. The two solid lines proportional to $e^{-0.905 ~\tau v/R}$ for $\tau v/R\le 1$, and $(\tau v/R)^{-1.101}$ for $\tau v/R \ge 1$ are plotted for comparison.
Parameters: $M=0.0565$, $\eta=1.5$, $R/L=0.06$.
}
\label{fig.var_Omega}
\end{center}
\end{figure}

To judge to what extent calculations with very small particle numbers predict the behavior of larger systems, 
we perform additional agent-based
simulations, see Figs.  
\ref{fig.SCv} and \ref{fig.SGv}.
These figures show how the correlation function scales as we increase the system's size and the particle number but 
keep the normalized density $M$ constant as well as the $\tau v/R$ ratio and the noise. 
We see a strong finite size effect altering the correlation functions. For large enough systems such as 
$N=100$ and $N=200$, the data is consistent with an initial power-law decay for the velocity correlation function $C_v$ 
followed by an exponential decay. This indicates that there exists a finite correlation length.
From Fig. \ref{fig.SCv} we read off a correlation length which is about an order of magnitude larger 
than both the interaction range $R$ and the mean free path $\lambda=\tau v_0$.
This is interesting because at $\eta=1$ we are deep into the disordered phase, quite far away from the onset of global collective motion.
This is consistent with the {\em precursor} phenomenon, reported in Ref. \cite{hanke_13}.

We also see that the correlation functions for different system sizes plotted as a function of $r/R$ roughly fall on  
top of each other, leading to a universal master curve. For small systems, the tail of the correlation function bends 
upward due to the boundary condition. The short distance behavior is then affected and therefore deviates from the master curve. 

In equilibrium spin systems at criticality, spin-spin correlations
decay with distance $r$ according to $\sim r^{-d+2-\eta}$, where $d$ is the spatial dimension and $\eta$
is a critical exponent which is usually quite small, $0\le \eta \le 0.25$.
Identifying spins with the velocity vectors ${\bf v}_i$ of self-propelled agents, analogies can be drawn.
For example, in a Vicsek-like 
system with an inner repulsion zone \cite{romenskyy_13a,lobaskin_14p}, an exponent of $\eta\approx 0.75$
was found right at the threshold to collective motion.
Cavagna {\em et al.} \cite{cavagna_10} investigated the velocity-velocity correlations 
inside three-dimensional flocks of starlings. 
These measurements correspond to the highly ordered regime, deep in the ordered phase.
They found a very weak decrease of the correlations, compatible either with a power law $\sim r^{-0.19}$,
a logarithmic decay, or even no decay, $\sim r^0$. 
In contrast, here, we are deep in the disordered phase, and 
the corresponding exponent $\approx 1.8$ 
shown in Fig. \ref{fig.SCv} is far from previously observed or anticipated values 
of $\eta$ at the transition point.
Note, that the results given by Toner, Tu and Ulm \cite{toner_95,tu_98} were mostly for density and velocity 
correlations in the strongly ordered regime, and thus cannot be related to our observations.
\begin{figure}[H]
\begin{center}
\includegraphics[width=3.in,angle=-90]{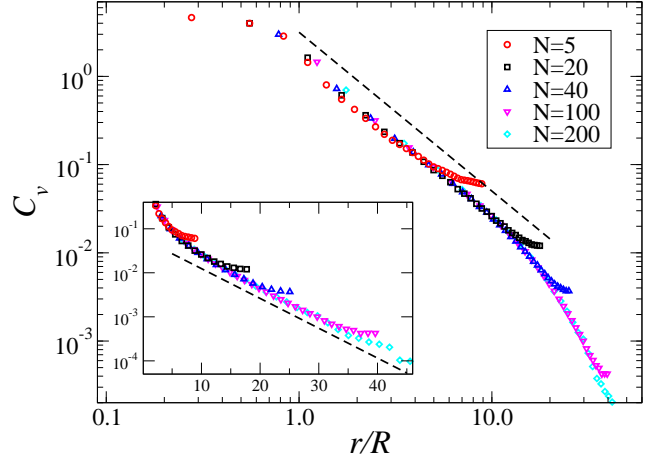}
\caption
{
The velocity correlation function for the agent-based simulations with various $N$.
The parameters $M=0.05$, $\tau v/R=1$ and $\eta=1.0$ are fixed for all the systems.
A power-law decay function (dashed line) with exponent $-1.8$ is plot for comparison.
The inset shows the same data but in log-normal scale with a dashed line proportional to $e^{-0.156 r/R}$.
}
\label{fig.SCv}
\end{center}
\end{figure}
\begin{figure}[H]
\begin{center}
\includegraphics[width=3.in,angle=-90]{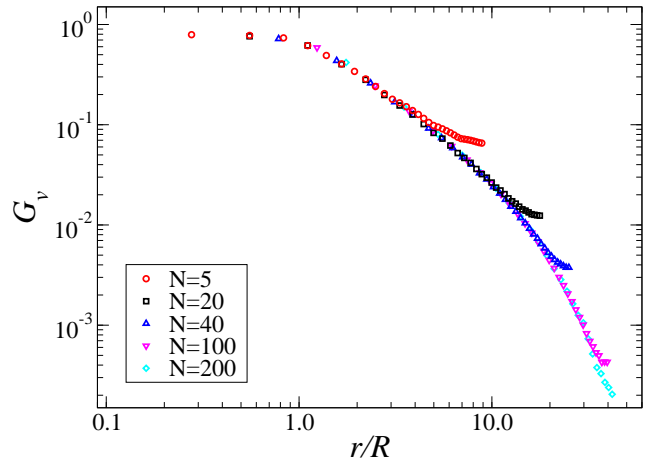}
\caption
{
Same as Fig. \ref{fig.SCv}, but this time that non-weighted velocity correlation function.
}
\label{fig.SGv}
\end{center}
\end{figure}
%
%
\section{Conclusion}
\label{sec:conclusion}

Very recently, it was discovered 
that correlation effects are not only important for 
a better quantitative description of active colloidal systems, but that they are {\em essential} for global phase ordering \cite{hanke_13,speck_14}.
It is likely that correlations play a similar important role in other experimental systems such as actin filaments \cite{schaller_10} or 
microtubules \cite{sumino_12} driven by molecular motors and vibrated
polar disks \cite{weber_13}.
So far, almost nothing specific is known
about correlations in active systems near the threshold to collective motion.
In this paper, we obtain orientational and spatial correlations from first principles for a Vicsek-style model.
This model serves as an archetype of active matter
and is easier to treat analytically than the experimental systems mentioned above.
In particular, we derive a repeated-ring kinetic theory for Vicsek-style models of self-propelled agents. 

The approach starts with an exact evolution equation for a Markov chain in phase space that incorporates the microscopic collision
rules. In contrast to our earlier approaches \cite{ihle_11,ihle_14_a,chou_12,romensky_14} and to most kinetic theories of active matter
we do not use the most severe approximation of kinetic theory -- the molecular chaos assumption. 
Instead of neglecting the connected two-particle correlations we derive 
an evolution equation for it: the second equation of a BBGKY-like hierarchy. 
Therefore, our theory goes beyond mean-field and is able to describe pre-collisional correlation as well as cluster formation in
a self-consistent way.
Both effects are important to correctly describe order/disorder transitions in Vicsek-style models at realistic physical parameters.
A correlated closure of the hierarchy is applied by 
neglecting
connected three- and higher multi-particle correlations.
By obtaining {\em quantitative agreement} between agent-based simulations and ring-kinetic predictions
for several correlation functions, we demonstrate 
that there is a weak-coupling regime in Vicsek-like models, where ring-kinetic theory gives correct results. This justifies the truncation 
of the BBGKY-hierarchy after the second equation in certain parameter ranges.

In order to facilitate the derivation of kinetic equations for self-propelled particle systems, we perform a 
small density expansion and introduce a novel diagrammatic technique to systematically account for terms in the collision integrals. 
We construct a Lattice-Boltzmann-like algorithm and numerically solve the ring-kinetic equations. The biggest difference to similar 
algorithms is that we propagate the two-particle correlation function in reduced space instead of merely 
dealing with the one-particle distribution.
We perform a detailed analysis of various density and orientational correlation functions by using both agent-based simulations and 
numerical solutions of ring-kinetic theory. Our results indicate significant pre-collisional correlations, unexpected oscillations 
and a quite large correlation length already in the disordered phase, quite far from the threshold to collective motion.
This could mean that, at least at small mean free paths, one might have to reinterpret the transition to collective motion in self-propelled particles 
as a transition from an orientationally 
correlated liquid to an even stronger correlated but ordered liquid \cite{hanke_13,weber_14}.
The observations of significant correlations in the disordered phase are consistent with the 
{\em precursor} phenomenon found in soft active colloids, \cite{hanke_13}.

Our results for the disordered phase are also reminiscent of recent 
experiments on swarms of midges \cite{attanasi_14} 
which show strong correlations despite a lack of
global order.
We found that the spatial behavior of the velocity correlation function is consistent with an initial power law decay with exponent $\approx -1.8$, followed
by an exponential decay. More research needs to be done to better understand this behavior. 
Using the diagrammatic kinetic formalism and the numerical results presented in this paper,
we hope to soon replace the numerical approach to the BBGKY equations by an analytical solution. 
This should allow us
to explore larger system sizes and to verify possible power-law regimes of the correlation functions.

We also discuss deviations between agent based simulations and ring-kinetic theory at very small noise and mean free path. 
One of the reasons for the discrepancies appears to be the existance of a strong-coupling regime 
where three-particle and higher multi-particle correlations dominate. Finding a suitable closure relation of the BBGKY hierarchy
for this case is related to the hardest problem of kinetic theory. This problem might be impossible 
to solve, and is left for future research.

The methods proposed in this paper could be extended to more realistic models of self-propelled particles, 
for example to the Vicsek-like model 
recently introduced by Lu {\em et al} \cite{lu_13} to explain
experiments
on the collective behavior of {\em Bacillus subtilis} in the presence of a photosensitizer.
Furthermore, our systematic derivation of correlation effects for a simplified model
could also be benefitial for calculations and an improved understanding of these effects in more complex experimental systems which cannot be
faithfully described by Vicsek-style models.
Finally, Vicsek-like models and models of granular matter
are somewhat similar 
with regard to the fact that the relative velocities of two particles are reduced during 
collisions by either
alignment or inelastic interactions, respectively.
Therefore, one can hope that the kinetic formalism for active matter proposed in this paper might also, in some way, become 
useful for granular 
matter.

\newpage
\appendix
\section{Coupling constants}
\label{app:coupling}

In this appendix, we give the integrals defined in Eq. (\ref{eq.coupling_definition}) of section \ref{sec.Fourier_expansion}.
For the standard Vicsek (VM) interaction rule and for arbitrary mode numbers, it is only possible to analytically calculate 
those coupling integrals which involve at most two particles per collision circle, for example $\mathrm{k}_{mpq}$, $\mathrm{j}_{mnpq}$, and $\mathrm{i}_{mnpqr}$. 
Apart from a few exceptions,
coupling integrals involving three or more particles per collision zone have to be evaluated numerically. 
This leads 
to intractable computational 
problems for large mode numbers. 
However, the binary Vicsek (BVM) interaction rule, 
where the focal particle randomly picks only one of their neighbors, allows us to break down the kernel of the integrand. 
For example,  in $\mathrm{k}_{mnpqr}$, the formal expression
 $e^{-im \Phi(\theta_1,\theta_2,\theta_3)}$ translates into $(e^{-im\Phi(\theta_1,\theta_2)}+e^{-im\Phi(\theta_1,\theta_3)})/2$ because
in BVM the focal particle (labeled 1) picks on of the two available particles $2$ and $3$ with equal probability $1/2$.

In this way one can write down the analytical form of coupling constants for all interactions with more than two particles per collision zone, 
provided that the basic units -- the binary couplings -- are given.
In Table \ref{ta.coupling}, we summarize the coupling integrals for both standard (VM) and binary Vicsek (BVM) interaction rules. 
\begin{table}[H]
\centering
\begin{tabular}{|c|c|c|c|}

\hline
diagram & coupling & standard Vicsek model &  binary Vicsek model \\
\hline
$\sCfF$ &                 
$\mathrm{k}_{mpq}$   & \multicolumn{2}{c|}{$\big\langle e^{-im \Phi(\theta_1,\theta_2)} \big\rangle$ }\\
\hline
$\sCfFF$ & 
$\mathrm{k}_{mpqr}$  & $\big\langle e^{-im \Phi(\theta_1,\theta_2,\theta_3)} \big\rangle$ 
                     & $\frac{1}{2}\big\langle e^{-im \Phi(\theta_1,\theta_2)}+e^{-im \Phi(\theta_1,\theta_3)} \big\rangle$\\
\hline
$\sDaff$ & 
$\mathrm{j}_{mnpq}$  & \multicolumn{2}{c|}{$\big\langle e^{-im \Phi(\theta_1,\theta_2)} e^{-in \Phi(\theta_1,\theta_2)}\big\rangle$ }\\
\hline
$\sDbffFk$ & 
$\mathrm{i}_{mnpqr}$ & \multicolumn{2}{c|}{$\big\langle e^{-im \Phi(\theta_1,\theta_3)} e^{-in \Phi(\theta_2,\theta_3)}\big\rangle$ }\\
\hline
$\sDaffFk$ & 
$\mathrm{h}_{mnpqr}$ & $\big\langle e^{-im \Phi(\theta_1,\theta_2,\theta_3)} e^{-in \Phi(\theta_1,\theta_2,\theta_3)}\big\rangle$ 
                     & $\frac{1}{4}\big\langle e^{-im \Phi(\theta_1,\theta_2)} e^{-in \Phi(\theta_2,\theta_1)}$\\ 
                   &&& $~+ e^{-im \Phi(\theta_1,\theta_2)} e^{-in \Phi(\theta_2,\theta_3)}$\\
                   &&& $~+ e^{-im \Phi(\theta_1,\theta_3)} e^{-in \Phi(\theta_2,\theta_1)}$\\
                   &&& $~+ e^{-im \Phi(\theta_1,\theta_3)} e^{-in \Phi(\theta_2,\theta_3)}\big\rangle$ \\  
\hline
$\sDaffFi$ & 
$\mathrm{l}_{mnpqr}$ & $\big\langle e^{-im \Phi(\theta_1,\theta_2,\theta_3)} e^{-in \Phi(\theta_1,\theta_2)}\big\rangle$ 
                     & $\frac{1}{2}\big\langle e^{-im \Phi(\theta_1,\theta_2)} e^{-in \Phi(\theta_2,\theta_1)}$ \\ 
                   &&& $~+ e^{-im \Phi(\theta_1,\theta_3)} e^{-in \Phi(\theta_2,\theta_1)}\big\rangle$ \\
\hline

\end{tabular}
\caption{The coupling constants for the VM and BVM. In this table $\big\langle\cdots\big\rangle$ means
$1/(2\pi)^2 \int d\theta_1 d\theta_2 \cdots e^{ip\theta_1}e^{iq\theta_2}$ for binary interaction, and
$1/(2\pi)^3 \int d\theta_1 d\theta_2 d\theta_3 \cdots e^{ip\theta_1}e^{iq\theta_2}e^{ir\theta_3}$ for 3-particle interaction.
The first column shows an example of diagram where the coupling constant applies to.
}
\label{ta.coupling}
\end{table}
Note, that the constant $h_{mnpqr}$ decomposes into four terms for the BVM. This is because the two focal particles have two possible 
choices each to pick a collision partner.

For binary collisions, the average angle is given by
\begin{equation}
\Phi(\theta_1,\theta_2) =
\begin{cases}
~~ \frac{\theta_1+\theta_2}{2}  & ~\text{for } 0\leq|\theta_1+\theta_2| < \pi\\
~~ \frac{\theta_1+\theta_2}{2}+\pi  & ~\text{otherwise } 
\end{cases}~.
\end{equation}
By switching the variables $\alpha=(\theta_1+\theta_2)/2$ and $\beta=(\theta_1-\theta_2)/2$,
the coupling $\mathrm{k}_{mpq}$ becomes
\begin{equation}
\mathrm{k}_{mpq}=\frac{1}{2\pi^2}\int_{-\pi}^{\pi}d\alpha\int_{-\pi/2}^{\pi/2} d\beta 
\,
{\rm e}^{-im\alpha} {\rm e}^{ip(\alpha+\beta )} {\rm e}^{iq(\alpha-\beta )},
\end{equation}
where the Jacobian, a factor of $2$, has been multiplied to the equation. We also changed the domain of the integration such that  $\Phi(\theta_1,\theta_2)=\alpha$ is continuous in the region and arrive at the following form 
\begin{equation}
\label{APP_K1}
\mathrm{k}_{mpq}=\frac{\sin[(m-p-q)\pi]}{(m-p-q)\pi}\frac{\sin[(p-q)\pi/2]}{(p-q)\pi/2}.
\end{equation}
We notice that the first factor is nothing but the Kronecker delta function $\delta_{m-p-q,0}$ since $m$, $p$, and $q$ are all integers.
Defining  
\begin{equation}
\mathrm{S}(x)\equiv \mathrm{sinc} \left(\frac{\pi x}{2}\right)=\frac{2}{\pi x}\sin\left(\frac{\pi x}{2}\right),
\end{equation}
Eq. (\ref{APP_K1}) becomes
\begin{equation}
\mathrm{k}_{mpq}= \mathrm{S}(p-q)\delta_{m,p+q}.
\label{eq.coupling_k1}
\end{equation}
The third coupling integral defined in Eq. (\ref{eq.coupling_definition}) of section \ref{sec.Fourier_expansion} is related to the first one 
by replacing $m$ by $m+n$ and can be written down immediately,
\begin{equation}
\mathrm{j}_{mnpq}= \mathrm{S}(p-q)\delta_{m+n,p+q}
\label{eq.coupling_j}
\end{equation}
The remaining coupling from Eq. (\ref{eq.coupling_definition}) 
that only involves two particles per circle,
the quantity $\mathrm{i}_{mnpqr}$, 
can be
calculated by realizing that the coupling integral $\mathrm{k}_{mpq}$ is actually the angular Fourier transform of
the factor $e^{-im\Phi(\theta_j,\theta_k)}$. Therefore, we plug
$e^{-im\Phi(\theta_j,\theta_k)}=\sum_{p,q}\mathrm{k}_{mpq}e^{-ip\theta_j}e^{-iq\theta_k}$
into the definition of the integral
\begin{eqnarray}
\mathrm{i}_{mnpqr}&=& \frac{1}{(2\pi)^3}\int d\theta_1 d\theta_2 d\theta_3~e^{i p \theta_1}e^{i q \theta_2}e^{i r \theta_3}
\\\nonumber
&&\times\left(\sum_{a,b}\mathrm{k}_{m a b}~e^{-i a\theta_1}~e^{-i b\theta_3}\right)
\left(\sum_{c,d}\mathrm{k}_{n c d}~e^{-i c\theta_2}~e^{-i d\theta_3}\right)
\\\nonumber
&=&\sum_{a,b,c,d}\mathrm{k}_{m a b}~\mathrm{k}_{n c d}~\delta_{p,a}~\delta_{q,c}~\delta_{r,b+d}
\\\nonumber
&=&\sum_{b}\mathrm{k}_{m, p, r-b}~\mathrm{k}_{n, q, b}
\end{eqnarray}
This way, $\mathrm{i}_{mnpqr}$ can be seen as a convolution of the coupling constant $\mathrm{k}_{mpq}$ with itself.

Using equation (\ref{eq.coupling_k1}), we have
\begin{equation}
\mathrm{i}_{mnpqr}= \mathrm{S}(m-2p)\mathrm{S}(n-2q)\delta_{m+n,p+q+r}
\label{eq.coupling_i}
\end{equation}
For the binary Vicsek model (BVM), all the other couplings can be derived from the three fundamental two-particle couplings (see Table \ref{ta.coupling})
\begin{eqnarray}
\mathrm{k}_{mpqr}&=& \frac{1}{2}\Big[ \mathrm{S}(p-q)\delta_{m,p+q}\delta_{r,0}+\mathrm{S}(p-r)\delta_{m,p+r}\delta_{q,0}\Big]
\\\nonumber
\mathrm{h}_{mnpqr}&=& \frac{1}{4}\Big[\mathrm{S}(p-q)\delta_{m+n,p+q}\delta_{r,0} \\\nonumber
  &&~ +\mathrm{S}(m-2p)\mathrm{S}(n-2q)\delta_{m+n,p+q+r} \\\nonumber
  &&~ +\mathrm{S}(m-2p)\mathrm{S}(n-2r)\delta_{m+n,p+q+r} \\\nonumber
  &&~ +\mathrm{S}(m-2r)\mathrm{S}(n-2q)\delta_{m+n,p+q+r} \Big]
\\\nonumber
\mathrm{l}_{mnpqr}&=& \frac{1}{2}\Big[ \mathrm{S}(p-q)\delta_{m+n,p+q}\delta_{r,0}+\mathrm{S}(m-2r)\mathrm{S}(n-2q)\delta_{m+n,p+q+r}\Big]~.
\end{eqnarray}
For the standard Vicsek interaction the quantity $\mathrm{k}_{mpqr}$ needs to be evaluated numerically.
Then, one can obtain $\mathrm{h}_{mnpqr}$ by the following relation
\begin{equation}
\label{A10}
\mathrm{h}_{mnpqr}=\mathrm{k}_{m+n,p,q,r}.
\end{equation}
The coupling $\mathrm{l}_{mnpqr}$ can be derived using the Fourier expansion of $e^{-im\Phi(\theta_j,\theta_k)}$ and of $e^{-im\Phi(\theta_j,\theta_k,\theta_l)}$ similarly to the way we derived the coupling $\mathrm{i}_{mnpqr}$ and arrive at
\begin{equation}
\label{A11}
\mathrm{l}_{mnpqr}=\sum_b ~\mathrm{k}_{m,p-b,q+b-n,r}~\mathrm{S}(n-2b).
\end{equation}
The result is further simplified to, 
\begin{equation}
\label{A12}
\mathrm{l}_{mnpqr}=
\begin{cases}
~~ \mathlarger{\sum}\limits_{\text{odd } b} ~\mathrm{k}_{m,p-\frac{n-b}{2},q-\frac{n+b}{2},r} ~\mathrm{S}(b)
&~\text{for odd } n\\
~~ \mathrm{k}_{m,p-\frac{n}{2},q-\frac{n}{2},r}  
& ~\text{for even } n
\end{cases}~.
\end{equation}
Note, that Eqs.(\ref{A10}-\ref{A12}) are general results that also apply to BVM.
\section{Diagrams for the second BBGKY-hierarchy equation}
\label{app:diagram}
Here, we consider contributions to the collision operator of the second hierarchy equation in Fourier space
as introduced in Section ~\ref{sec:diagram}. 
The complete list of terms for a low density expansion to order
$O(M^3)$ in diagrammatic form is: 

{\bf Strong overlap}
\begin{eqnarray}
\blangle \sDaff \brangle_{mn} &=& 
\lambda_{mn}\sum_{pq}\mathrm{j}_{mnpq}
\hat{f}_p(\mathbf{x}_1)\hat{f}_q(\mathbf{x}_2)
\\
\blangle \sDag \brangle_{mn} &=& 
\lambda_{mn}\sum_{pq}\mathrm{j}_{mnpq}
\hat{g}_{pq}(\mathbf{x}_1,\mathbf{x}_2)
\end{eqnarray}
\begin{eqnarray}
\blangle\sDaffFi\brangle_{mn}&=& 
2\pi\lambda_{mn}\sum_{pqr}\mathrm{l}_{mnpqr}~
\hat{f}_p(\mathbf{x}_1) 
\hat{f}_q(\mathbf{x}_2) 
\Big[
\bar{F}_r(\mathbf{x}_1)
-\Delta\bar{F}_r(\mathbf{x}_1,\mathbf{x}_2)
\Big]
\\
\blangle\sDaffFj\brangle_{mn}&=& 
2\pi\lambda_{mn}\sum_{pqr}\mathrm{l}_{nmqpr}~
\hat{f}_q(\mathbf{x}_2) 
\hat{f}_p(\mathbf{x}_1)
\Big[ 
\bar{F}_r(\mathbf{x}_2)
-\Delta\bar{F}_r(\mathbf{x}_2,\mathbf{x}_1)
\Big]
\\
\blangle\sDaffFk\brangle_{mn}&=& 
2\pi\lambda_{mn}\sum_{pqr}\mathrm{h}_{mnpqr}~
\hat{f}_p(\mathbf{x}_1) 
\hat{f}_q(\mathbf{x}_2) 
\Delta\bar{F}_r(\mathbf{x}_1,\mathbf{x}_2)
\end{eqnarray}
\begin{eqnarray}
\blangle\sDagFi\brangle_{mn}&=& 
2\pi\lambda_{mn}\sum_{pqr}\mathrm{l}_{mnpqr}~
\hat{g}_{pq}(\mathbf{x}_1,\mathbf{x}_2) 
\Big[
\bar{F}_r(\mathbf{x}_1)
-\Delta\bar{F}_r(\mathbf{x}_1,\mathbf{x}_2)
\Big]
\\
\blangle\sDagFj\brangle_{mn}&=& 
2\pi\lambda_{mn}\sum_{pqr}\mathrm{l}_{nmqpr}~
\hat{g}_{qp}(\mathbf{x}_2,\mathbf{x}_1) 
\Big[
\bar{F}_r(\mathbf{x}_2)
-\Delta\bar{F}_r(\mathbf{x}_2,\mathbf{x}_1)
\Big]
\\
\blangle\sDagFk\brangle_{mn}&=& 
2\pi\lambda_{mn}\sum_{pqr}\mathrm{h}_{mnpqr}~
\hat{g}_{pq}(\mathbf{x}_1,\mathbf{x}_2) 
\Delta\bar{F}_r(\mathbf{x}_1,\mathbf{x}_2)
\end{eqnarray}
\begin{eqnarray}
\blangle\sDaGif\brangle_{mn}&=& 
2\pi\lambda_{mn}\sum_{pqr}\mathrm{l}_{mnpqr}
\Big[
\bar{G}_{pr}(\mathbf{x}_1,\mathbf{x}_1) 
-\Delta\bar{G}_{pr}(\mathbf{x}_1,\mathbf{x}_2) 
\Big]
\hat{f}_q(\mathbf{x}_2)
\\
\blangle\sDaGjf\brangle_{mn}&=& 
2\pi\lambda_{mn}\sum_{pqr}\mathrm{l}_{nmqpr}
\Big[
\bar{G}_{pr}(\mathbf{x}_1,\mathbf{x}_2) 
-\Delta\bar{G}_{pr}(\mathbf{x}_1,\mathbf{x}_2) 
\Big]
\hat{f}_q(\mathbf{x}_2)
\\
\blangle\sDaGkf\brangle_{mn}&=& 
2\pi\lambda_{mn}\sum_{pqr}\mathrm{h}_{mnpqr}~
\Delta\bar{G}_{pr}(\mathbf{x}_1,\mathbf{x}_2) 
\hat{f}_q(\mathbf{x}_2)
\end{eqnarray}
\begin{eqnarray}
\blangle\sDafGi\brangle_{mn}&=&
2\pi\lambda_{mn}\sum_{pqr}\mathrm{l}_{mnpqr}~
\hat{f}_p(\mathbf{x}_1)
\Big[ 
\bar{G}_{qr}(\mathbf{x}_2,\mathbf{x}_1) 
-\Delta\bar{G}_{qr}(\mathbf{x}_2,\mathbf{x}_1) 
\Big]
\\
\blangle\sDafGj\brangle_{mn}&=&
2\pi\lambda_{mn}\sum_{pqr}\mathrm{l}_{nmqpr}~
\hat{f}_p(\mathbf{x}_1)
\Big[
\bar{G}_{qr}(\mathbf{x}_2,\mathbf{x}_2) 
-\Delta\bar{G}_{qr}(\mathbf{x}_2,\mathbf{x}_1) 
\Big]
\\
\blangle\sDafGk\brangle_{mn}&=&
2\pi\lambda_{mn}\sum_{pqr}\mathrm{h}_{mnpqr}~
\hat{f}_p(\mathbf{x}_1)
\Delta\bar{G}_{qr}(\mathbf{x}_2,\mathbf{x}_1) 
\end{eqnarray}
\begin{eqnarray}
\blangle\sDaHif\brangle_{mn}&=& 
-2\pi\lambda_{mn}\sum_{pq}\mathrm{j}_{mnpq}
\Big[
\bar{G}_{p0}(\mathbf{x}_1,\mathbf{x}_1) 
-\Delta\bar{G}_{p0}(\mathbf{x}_1,\mathbf{x}_2) 
\Big]
\hat{f}_q(\mathbf{x}_2)
\\
\blangle\sDaHjf\brangle_{mn}&=& 
-2\pi\lambda_{mn}\sum_{pq}\mathrm{j}_{mnpq}
\Big[
\bar{G}_{p0}(\mathbf{x}_1,\mathbf{x}_2) 
-\Delta\bar{G}_{p0}(\mathbf{x}_1,\mathbf{x}_2) 
\Big]
\hat{f}_q(\mathbf{x}_2)
\\
\blangle\sDaHkf\brangle_{mn}&=& 
-2\pi\lambda_{mn}\sum_{pq}\mathrm{j}_{mnpq}~
\Delta\bar{G}_{p0}(\mathbf{x}_1,\mathbf{x}_2) 
\hat{f}_q(\mathbf{x}_2)
\end{eqnarray}
\begin{eqnarray}
\blangle\sDafHi\brangle_{mn}&=&
-2\pi\lambda_{mn}\sum_{pq}\mathrm{j}_{mnpq}~
\hat{f}_p(\mathbf{x}_1)
\Big[ 
\bar{G}_{q0}(\mathbf{x}_2,\mathbf{x}_1) 
-\Delta\bar{G}_{q0}(\mathbf{x}_2,\mathbf{x}_1) 
\Big]
\\
\blangle\sDafHj\brangle_{mn}&=&
-2\pi\lambda_{mn}\sum_{pq}\mathrm{j}_{mnpq}~
\hat{f}_p(\mathbf{x}_1)
\Big[
\bar{G}_{q0}(\mathbf{x}_2,\mathbf{x}_2) 
-\Delta\bar{G}_{q0}(\mathbf{x}_2,\mathbf{x}_1) 
\Big]
\\
\blangle\sDafHk\brangle_{mn}&=&
-2\pi\lambda_{mn}\sum_{pq}\mathrm{j}_{mnpq}~
\hat{f}_p(\mathbf{x}_1)
\Delta\bar{G}_{q0}(\mathbf{x}_2,\mathbf{x}_1) 
\end{eqnarray}

{\bf Weak overlap}
\begin{eqnarray}
\blangle \sDbff \brangle_{mn} &=& 
\lambda_{mn}
\hat{f}_m(\mathbf{x}_1)\hat{f}_n(\mathbf{x}_2)
\\
\blangle \sDbg \brangle_{mn} &=& 
\lambda_{mn}
\hat{g}_{mn}(\mathbf{x}_1,\mathbf{x}_2)
\end{eqnarray}
\begin{eqnarray}
\blangle\sDbffFi\brangle_{mn}&=& 
2\pi\lambda_{mn}\sum_{pr}\mathrm{k}_{mpr}~
\hat{f}_p(\mathbf{x}_1) 
\Big[
\bar{F}_r(\mathbf{x}_1)
-\Delta\bar{F}_r(\mathbf{x}_1,\mathbf{x}_2)
\Big]
\hat{f}_n(\mathbf{x}_2) 
\\
\blangle\sDbffFj\brangle_{mn}&=& 
2\pi\lambda_{mn}\sum_{qr}\mathrm{k}_{nqr}~
\hat{f}_m(\mathbf{x}_1)
\hat{f}_q(\mathbf{x}_2) 
\Big[ 
\bar{F}_r(\mathbf{x}_2)
-\Delta\bar{F}_r(\mathbf{x}_2,\mathbf{x}_1)
\Big]
\\
\blangle\sDbffFk\brangle_{mn}&=& 
2\pi\lambda_{mn}\sum_{pqr}\mathrm{i}_{mnpqr}~
\hat{f}_p(\mathbf{x}_1) 
\hat{f}_q(\mathbf{x}_2) 
\Delta\bar{F}_r(\mathbf{x}_1,\mathbf{x}_2)
\end{eqnarray}
\begin{eqnarray}
\blangle\sDbgFi\brangle_{mn}&=& 
2\pi\lambda_{mn}\sum_{pr}\mathrm{k}_{mpr}~
\hat{g}_{pn}(\mathbf{x}_1,\mathbf{x}_2) 
\Big[
\bar{F}_r(\mathbf{x}_1)
-\Delta\bar{F}_r(\mathbf{x}_1,\mathbf{x}_2)
\Big]
\\
\blangle\sDbgFj\brangle_{mn}&=& 
2\pi\lambda_{mn}\sum_{qr}\mathrm{k}_{nqr}~
\hat{g}_{mq}(\mathbf{x}_1,\mathbf{x}_2) 
\Big[
\bar{F}_r(\mathbf{x}_2)
-\Delta\bar{F}_r(\mathbf{x}_2,\mathbf{x}_1)
\Big]
\\
\blangle\sDbgFk\brangle_{mn}&=& 
2\pi\lambda_{mn}\sum_{pqr}\mathrm{i}_{mnpqr}~
\hat{g}_{pq}(\mathbf{x}_1,\mathbf{x}_2) 
\Delta\bar{F}_r(\mathbf{x}_1,\mathbf{x}_2)
\end{eqnarray}
\begin{eqnarray}
\blangle\sDbGif\brangle_{mn}&=& 
2\pi\lambda_{mn}\sum_{pr}\mathrm{k}_{mpr}
\Big[
\bar{G}_{pr}(\mathbf{x}_1,\mathbf{x}_1) 
-\Delta\bar{G}_{pr}(\mathbf{x}_1,\mathbf{x}_2) 
\Big]
\hat{f}_n(\mathbf{x}_2)
\\
\blangle\sDbGjf\brangle_{mn}&=& 
2\pi\lambda_{mn}\sum_{qr}\mathrm{k}_{nqr}
\Big[
\bar{G}_{mr}(\mathbf{x}_1,\mathbf{x}_2) 
-\Delta\bar{G}_{mr}(\mathbf{x}_1,\mathbf{x}_2) 
\Big]
\hat{f}_q(\mathbf{x}_2)
\\
\blangle\sDbGkf\brangle_{mn}&=& 
2\pi\lambda_{mn}\sum_{pqr}\mathrm{i}_{mnpqr}~
\Delta\bar{G}_{pr}(\mathbf{x}_1,\mathbf{x}_2) 
\hat{f}_q(\mathbf{x}_2)
\end{eqnarray}
\begin{eqnarray}
\blangle\sDbfGi\brangle_{mn}&=&
2\pi\lambda_{mn}\sum_{pr}\mathrm{k}_{mpr}~
\hat{f}_p(\mathbf{x}_1)
\Big[ 
\bar{G}_{nr}(\mathbf{x}_2,\mathbf{x}_1) 
-\Delta\bar{G}_{nr}(\mathbf{x}_2,\mathbf{x}_1) 
\Big]
\\
\blangle\sDbfGj\brangle_{mn}&=&
2\pi\lambda_{mn}\sum_{qr}\mathrm{k}_{nqr}~
\hat{f}_m(\mathbf{x}_1)
\Big[
\bar{G}_{qr}(\mathbf{x}_2,\mathbf{x}_2) 
-\Delta\bar{G}_{qr}(\mathbf{x}_2,\mathbf{x}_1) 
\Big]
\\
\blangle\sDbfGk\brangle_{mn}&=&
2\pi\lambda_{mn}\sum_{pqr}\mathrm{i}_{mnpqr}~
\hat{f}_p(\mathbf{x}_1)
\Delta\bar{G}_{qr}(\mathbf{x}_2,\mathbf{x}_1) 
\end{eqnarray}
\begin{eqnarray}
\blangle\sDbHif\brangle_{mn}&=& 
-2\pi\lambda_{mn}
\Big[
\bar{G}_{m0}(\mathbf{x}_1,\mathbf{x}_1) 
-\Delta\bar{G}_{m0}(\mathbf{x}_1,\mathbf{x}_2) 
\Big]
\hat{f}_n(\mathbf{x}_2)
\\
\blangle\sDbHjf\brangle_{mn}&=& 
-2\pi\lambda_{mn}
\Big[
\bar{G}_{m0}(\mathbf{x}_1,\mathbf{x}_2) 
-\Delta\bar{G}_{m0}(\mathbf{x}_1,\mathbf{x}_2) 
\Big]
\hat{f}_n(\mathbf{x}_2)
\\
\blangle\sDbHkf\brangle_{mn}&=& 
-2\pi\lambda_{mn}
\Delta\bar{G}_{m0}(\mathbf{x}_1,\mathbf{x}_2) 
\hat{f}_n(\mathbf{x}_2)
\end{eqnarray}
\begin{eqnarray}
\blangle\sDbfHi\brangle_{mn}&=&
-2\pi\lambda_{mn}
\hat{f}_m(\mathbf{x}_1)
\Big[ 
\bar{G}_{n0}(\mathbf{x}_2,\mathbf{x}_1) 
-\Delta\bar{G}_{n0}(\mathbf{x}_2,\mathbf{x}_1) 
\Big]
\\
\blangle\sDbfHj\brangle_{mn}&=&
-2\pi\lambda_{mn}
\hat{f}_m(\mathbf{x}_1)
\Big[
\bar{G}_{n0}(\mathbf{x}_2,\mathbf{x}_2) 
-\Delta\bar{G}_{n0}(\mathbf{x}_2,\mathbf{x}_1) 
\Big]
\\
\blangle\sDbfHk\brangle_{mn}&=&
-2\pi\lambda_{mn}
\hat{f}_m(\mathbf{x}_1)
\Delta\bar{G}_{n0}(\mathbf{x}_2,\mathbf{x}_1) 
\end{eqnarray}
%
%

{\bf No overlap}
\begin{eqnarray}
\blangle \sDcff \brangle_{mn} &=& 
\lambda_{mn}
\hat{f}_m(\mathbf{x}_1)\hat{f}_n(\mathbf{x}_2)
\\
\blangle \sDcg \brangle_{mn} &=& 
\lambda_{mn}
\hat{g}_{mn}(\mathbf{x}_1,\mathbf{x}_2)
\end{eqnarray}
\begin{eqnarray}
\blangle\sDcffFi\brangle_{mn}&=& 
2\pi\lambda_{mn}\sum_{pr}\mathrm{k}_{mpr}~
\hat{f}_p(\mathbf{x}_1) 
\bar{F}_r(\mathbf{x}_1)
\hat{f}_n(\mathbf{x}_2) 
\\
\blangle\sDcffFj\brangle_{mn}&=& 
2\pi\lambda_{mn}\sum_{qr}\mathrm{k}_{nqr}~
\hat{f}_m(\mathbf{x}_1)
\hat{f}_q(\mathbf{x}_2) 
\bar{F}_r(\mathbf{x}_2)
\end{eqnarray}
\begin{eqnarray}
\blangle\sDcgFi\brangle_{mn}&=& 
2\pi\lambda_{mn}\sum_{pr}\mathrm{k}_{mpr}~
\hat{g}_{pn}(\mathbf{x}_1,\mathbf{x}_2) 
\bar{F}_r(\mathbf{x}_1)
\\
\blangle\sDcgFj\brangle_{mn}&=& 
2\pi\lambda_{mn}\sum_{qr}\mathrm{k}_{nqr}~
\hat{g}_{mq}(\mathbf{x}_1,\mathbf{x}_2) 
\bar{F}_r(\mathbf{x}_2)
\end{eqnarray}
\begin{eqnarray}
\blangle\sDcGif\brangle_{mn}&=& 
2\pi\lambda_{mn}\sum_{pr}\mathrm{k}_{mpr}
\bar{G}_{pr}(\mathbf{x}_1,\mathbf{x}_1) 
\hat{f}_n(\mathbf{x}_2)
\\
\blangle\sDcGjf\brangle_{mn}&=& 
2\pi\lambda_{mn}\sum_{qr}\mathrm{k}_{nqr}
\bar{G}_{mr}(\mathbf{x}_1,\mathbf{x}_2) 
\hat{f}_q(\mathbf{x}_2)
\end{eqnarray}
\begin{eqnarray}
\blangle\sDcfGi\brangle_{mn}&=&
2\pi\lambda_{mn}\sum_{pr}\mathrm{k}_{mpr}~
\hat{f}_p(\mathbf{x}_1)
\bar{G}_{nr}(\mathbf{x}_2,\mathbf{x}_1) 
\\
\blangle\sDcfGj\brangle_{mn}&=&
2\pi\lambda_{mn}\sum_{qr}\mathrm{k}_{nqr}~
\hat{f}_m(\mathbf{x}_1)
\bar{G}_{qr}(\mathbf{x}_2,\mathbf{x}_2) 
\end{eqnarray}
\begin{eqnarray}
\blangle\sDcHif\brangle_{mn}&=& 
-2\pi\lambda_{mn}
\bar{G}_{m0}(\mathbf{x}_1,\mathbf{x}_1) 
\hat{f}_n(\mathbf{x}_2)
\\
\blangle\sDcHjf\brangle_{mn}&=& 
-2\pi\lambda_{mn}
\bar{G}_{m0}(\mathbf{x}_1,\mathbf{x}_2) 
\hat{f}_n(\mathbf{x}_2)
\end{eqnarray}
\begin{eqnarray}
\blangle\sDcfHi\brangle_{mn}&=&
-2\pi\lambda_{mn}
\hat{f}_m(\mathbf{x}_1)
\bar{G}_{n0}(\mathbf{x}_2,\mathbf{x}_1) 
\\
\blangle\sDcfHj\brangle_{mn}&=&
-2\pi\lambda_{mn}
\hat{f}_m(\mathbf{x}_1)
\bar{G}_{n0}(\mathbf{x}_2,\mathbf{x}_2) 
\end{eqnarray}

\section{Limitations of the ring-kinetic approach}
In this Appendix we show some results with discrepancies
between theory and agent-based simulations. Some deviations have a simple numerical origin
and could be remedified by using more CPU time and memory. Others are due to the fundamental limitations
of a low density expansion or the neglect of connected three-particle and higher multi-particle correlations. 
We notice that in some cases there might be significant errors in 
the density correlation $C_\rho$ and the velocity correlation $C_v$, whereas
the agreement for the non-weighted velocity correlation $G_v$ is still very good, see for example Fig \ref{fig.a2}. 
Therefore, to discuss the limitations of the theory one has to carefully inspect all three quantities, $C_\rho$, $C_v$ and $G_v$.
\begin{figure}[H]
\begin{center}
\includegraphics[width=6.in]{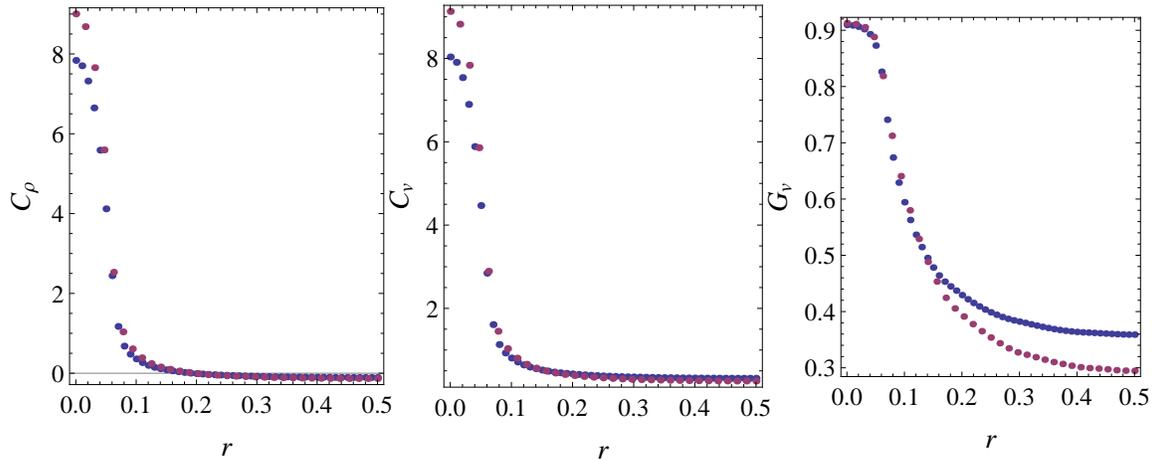}
\caption
{
Correlation functions for a $5$-particle system with $\eta=0.6$, $M=0.0565$, $R/L=0.06$ and $\tau v/R=1$.
The blue dots show the numerical evaluation of the kinetic theory, red stands for agent-based simulations.
The system size is 100 lattice units.
}
\label{fig.a1}
\end{center}
\end{figure}
\begin{figure}[H]
\begin{center}
\includegraphics[width=6.in]{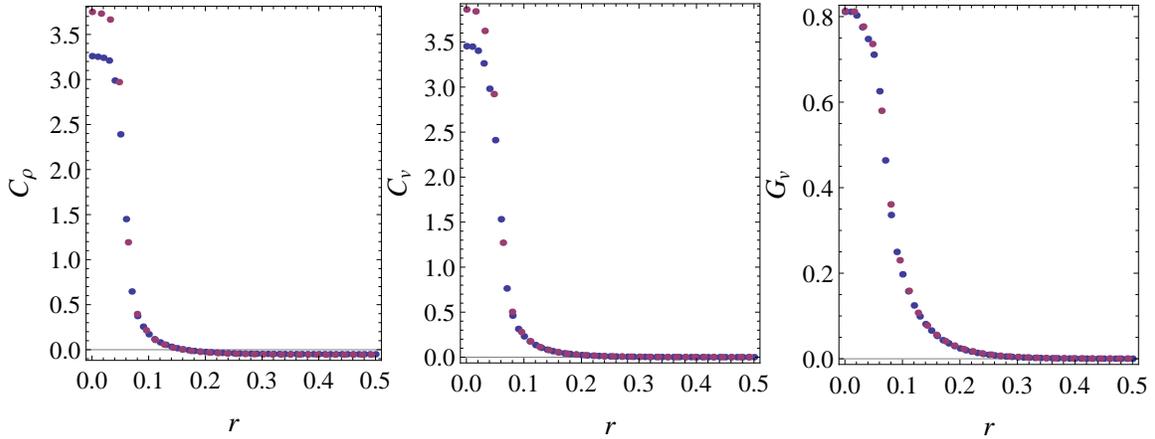}
\caption
{
Correlation functions for a $5$-particle system with $\tau v/R=1/3$, $\eta=1.5$, $M=0.0565$, and $R/L=0.06$.
The blue dots show the numerical result for theory and red for agent-based simulations.
The system size is 100 lattice units.
}
\label{fig.a2}
\end{center}
\end{figure}
\begin{figure}[H]
\begin{center}
\includegraphics[width=6.in]{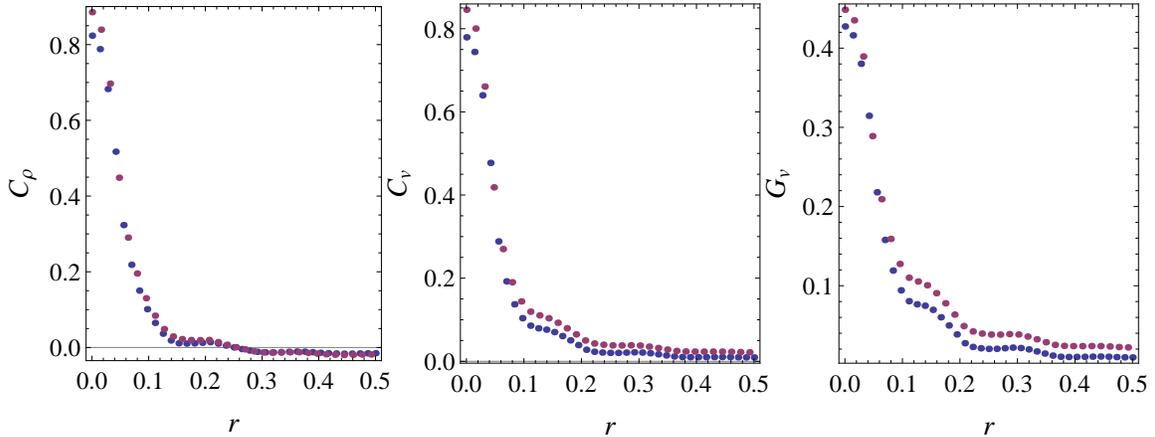}
\caption
{
Correlation functions for a $40$-particle system with $M=0.2182$, $R/L=0.0417$, $\eta=1.5$ and $\tau v/R=2$.
The blue dots show the numerical result for theory and red for agent-based simulation.
The system size is 72 lattice units but rescaled to $L=1$ in the plot.
}
\label{fig.a3}
\end{center}
\end{figure}
%
%
In Fig. \ref{fig.a1} we take the parameters of the $5$-particle system shown before in Fig. \ref{fig.Gv-N5} (cyan squares)
and reduce the noise from $1.5$ to $\eta=0.6$.
We observe that the kinetic theory now overestimates the value of $G_v$ outside the collision zone by up to $25\%$.
Furthermore, we see that the agent-based simulations (red dots) give larger values for both $C_{\rho}$ and $C_v$ near the center of
the collision circle. 

Comparing Fig. \ref{fig.Gv-N5} with $G_v$ from Fig. \ref{fig.a1}, it is clear that pre-collisional correlations are now stronger.
This is because 
the smaller noise 
makes particles stay together longer after a collision. 
According to the discussion in Ref. \cite{ihle_14_a}, it is fair to assume that also the three- four- and five-particle correlations have gained in strength. Therefore, a plausible source of the discrepancy in Fig. \ref{fig.a1} is 
the neglect of these higher multi-particle correlations in our theory. 
Note, that to rule out another reason for deviations, for this calculation we truncated 
the angular Fourier modes after the 21st mode instead of the typical truncation after $\pm 11$ modes.
This is because, on average, particles come out of a collision with directions inside an angular cone of width $\eta$.
For small noise this corresponds to a
rather sharp peak in angular space. To resolve it, at least approximately $2\pi/\eta$ modes are needed.
For $\eta=0.6$ this gives $11$ as minimum mode number which is much lower than the $21$ we used here.
Note that, currently, solving both BBGKY-equations simultaneously and 
lowering the noise to values around the transition threshold for 
collective motion is not feasible due to numerical instabilities.
A possible reason is that for the low densities $M\ll 1$ our kinetic approach is restricted to, 
the critical noise $\eta_C\sim \sqrt{M}$, is too small to
be represented by 21 Fourier modes. Work to extend the approach to larger density is underway \cite{ihle_14_proceeding}. 

To investigate the effects of small mean free path, starting again from the 5-particle system of Fig. \ref{fig.Gv-N5}, 
we reduce the mean free path ratio $\lambda/R$ from 
$2/3$ to $1/3$. Fig. \ref{fig.a2} shows that while there is no discrepancies in $G_v$, the theory underestimates $C_v$ and $C_{\rho}$
at small distances by up to $15\%$.
In Ref. \cite{ihle_14_a} it was shown that at small mean free path, clustering becomes strong. 
That is, even at very small densities, $M\ll 1$,
there is a large likelihood to find more than two particles in a collision circle.
Thus, again, a likely 
source of the deviations is that the kinetic theory neglects 
higher multi-particle correlations. 
Another possible source of the devations is that 
at small mean free path ratios $\lambda/R$,
the mean free path is usually discretized by only a few lattice units, in this case by only $2$ lattice units.
In other tests (not shown) we observed discretization errors when, depending on noise strength, $\lambda$ was 
discretized by less than 3 to 4 lattice units.

Finally, in Fig. \ref{fig.a3}
we explore the limits of the low density expansion and study a system with $M=0.2182$.
The small discrepancies in all three functions $C_v$, $C_{\rho}$ and $G_v$ look qualitatively different than in Figs. \ref{fig.a1} and 
\ref{fig.a2}, and are likeley caused by neglecting diagrams of higher order than
$O(M^3)$ in our diagrammatic expansion.


\begin{thebibliography}{99}
\bibitem{FOOT0}
By ``moderate clustering'' it is meant that the cluster formation is still weak enough to be describable 
by the two-particle connected
correlation functions and that higher multi-particle correlations are negligible.
\bibitem{FOOT1}
In contrast to regular gases, a large number of collision partners is possible here because
the particles in the VM have zero volume.
\bibitem{FOOT2}
assuming non-zero noise $\eta>0$ and a system size $L$ diverging so strongly  
that also $L/\lambda\rightarrow \infty$.
\bibitem{FOOT3}
This is because Eq. (\ref{BINOMIAL_F}) is a polynomial in $x=M/N$ that sums up to unity
if all terms were included. For this to be possible, terms of each order in $x^S$ must cancel each other.
Hence, terms linear in $x$ will cancel, all quadratic terms will cancel, and so on.
\bibitem{FOOT4}
Note, that according to the discussion in Section ~\ref{sec:verify}, the 
condition $\langle \Omega \rangle=0$ 
does not necessarily restrict us to the disordered phase.
This is because the kinetic theory approach is based on ensemble-averaging. The quantities $f_1$, $f_2$ and $g_2$ result from 
such an average. Depending on the chosen members of
the ensemble, $\langle \Omega \rangle$ could well be zero at parameter ranges where agent-based simulations show a strongly ordered state.
This would be the case if realizations of the system with all possible directions of the (long) total momentum vector would be averaged over.
\end{thebibliography}
\end{document}